\documentclass[12pt]{article}

\usepackage{latexsym,amsmath,amssymb,amsthm,amsfonts,graphicx,natbib,breqn,epsfig,bm,authblk,url,thmtools}
\usepackage{algorithm}

\usepackage{float} 
\usepackage{booktabs} 
\usepackage[margin=1cm]{caption} 
\usepackage{mathtools}
\usepackage{enumerate}
\usepackage{bbm}
\usepackage{tikz}
\usepackage[colorinlistoftodos]{todonotes}
\usetikzlibrary{shapes.multipart}
\usetikzlibrary{arrows, positioning}

\floatstyle{plain}
\newfloat{Algorithm}{thp}{lop}
\floatname{Algorithm}{Algorithm}

\declaretheoremstyle[
notefont=\bfseries, notebraces={}{},
bodyfont=\normalfont,
postheadspace=0.5em,
numbered=yes,
]{mystyle}

\theoremstyle{definition}

\def\yt{\widetilde{y}}

\def\yc{\check{y}}

\newcommand{\blind}{0}

\begin{document}

\def\spacingset#1{\renewcommand{\baselinestretch}%
	{#1}\small\normalsize} \spacingset{1}

\if0\blind
{
	\title{\bf Bayesian clustering using random effects models and predictive projections}
	\author[1,2]{Yinan Mao\thanks{
		The author gratefully acknowledges the funding support from the Singapore Population Health Improvement Centre (SPHERiC).}}
	\author[1,3]{David J. Nott}
	\affil[1]{Department of Statistics and Applied Probability, National University of Singapore, Singapore 117546}
	\affil[2]{Saw Swee Hock School Of Public Health, National University of Singapore, Singapore 117549}
	\affil[3]{Operations Research and Analytics Cluster, National University of Singapore, Singapore 119077}
	\maketitle
} \fi

\if1\blind
{
	\bigskip
	\bigskip
	\bigskip
	\begin{center}
		{\LARGE\bf Bayesian clustering using random effects models and predictive projections}
	\end{center}
	\medskip
} \fi

%\bigskip
\begin{abstract}
	Linear mixed models are widely used for analyzing 
	hierarchically structured data involving missingness and unbalanced study designs.  
	We consider a Bayesian clustering method
	that combines linear mixed models and predictive projections.  
	For each observation, we consider a predictive replicate
	in which only a subset of the random effects is shared between the observation and its replicate,
	with the remainder being integrated out using the conditional prior.  Predictive projections are then defined 
	in which the number of distinct values taken by the shared random effects is finite, in order to obtain different 
	clusters. Integrating out some of the random effects acts as a noise filter, allowing the clustering to be
	focused on only certain chosen features of the data.    The method
	is inspired by methods for Bayesian model checking, in which 
	simulated data replicates from a fitted model are used for model criticism by
	examining their similarity to the observed data in relevant ways.  Here
	the predictive replicates are used to define similarity between observations in relevant ways for clustering.
	To illustrate the way our method reveals
	aspects of the data at different scales,  we consider fitting temporal
	trends in longitudinal data using Fourier cosine bases with a random
	effect for each basis function, and different clusterings defined by shared random effects for replicates
	of low or high frequency
	terms.  The method is demonstrated in a series of real examples.
\end{abstract}

\noindent
{\it Keywords:}  Bayesian clustering, linear mixed models, longitudinal data, predictive projections.\\
%{\it Word count:} 7024 
\vfill

\newpage
\spacingset{1.5} % DON'T change the spacing!

\section{Introduction}\label{sec:Intro}

Linear mixed models are widely used for analyzing longitudinal and other hierarchically structured 
data involving unbalanced designs or missingness and correlations between observations.  For cluster analysis 
of complex longitudinal datasets, many authors have considered mixture and partition models with linear mixed model components.  
These models provide attractive model-based approaches to clustering in many problems.  However, computational aspects of these
methods are challenging, and choosing the number of clusters by conventional model choice criteria does not account
for what the clustering will be used for.  Here we consider an alternative Bayesian approach for 
model-based clustering using linear mixed models which does not need to use mixture or partition models, although
it can.   After fitting a linear mixed model, the method
considers a predictive replicate for each observation, in which some random effects are shared
with the original observation and the remaining random effects are integrated out using the conditional prior.
For the resulting predictive distributions, we consider predictive projections in which
the number of distinct values for the shared random effects is finite, defining different clusters.  
Our method extends predictive projection
approaches for variable selection \citep{dupuis_variable_2003,piironen+pv20} to clustering.  

The main advantage of the method is the ability
to control what aspects of the data define clusters through the choice
of random effects which are shared with the predictive replicates.  
Ignoring some information in order to simplify is an important part of any clustering method, and
integrating out some of the random effects via the conditional prior 
makes the choice of what should be ignored explicit.  
The consideration of predictive replicates with various kinds of replication 
is common in Bayesian model checking \citep{gelman+ms96}, where they
are used to judge whether replicated data from a fitted model ``look like'' the observed data in relevant ways.  
We make a related use of predictive replicates here, where the predictive distributions for replicates are used in
clustering to judge whether different observations are similar.  
As an example, later we consider fitting temporal trends in longitudinal data using a 
Fourier cosine basis, and different clusterings resulting from choosing the shared
random effects between the original observations and replicates 
as the high or low frequency terms.  The different clusterings are able
to reveal structure at different temporal scales.

Mixtures of linear mixed models are perhaps the most natural way to extend usual parametric
mixture models \citep{mclachlan+p00,bouveyron+cbr19}
to the clustering of longitudinal data with complex structure. Two early papers following this approach are
\cite{bar-joseph+ggjs02} and \cite{luan+l03}, who considered mixtures of mixed effects models 
with cubic spline and B-spline basis expansions respectively.  
\cite{pfeifer04} considers clustering based on a mixed model with a normal mixture model for the random effects.  
Mixtures of linear mixed models with gene level random effects 
in gene expression studies with replicates were considered in 
\cite{celeux+ml05}.  Their work was extended by   
\cite{ng+mwjn06}, who considered a general framework
with random effects at both the gene and tissue level. 
\cite{ray+m06} consider a multi-scale approach using a wavelet basis and a Dirichlet process
prior for the curve specific parameters.  \cite{coke+t10} consider random effects
mixture models with flexible time series structure based on antedependence models.
Computation with mixtures of linear mixed models is difficult, and \cite{scharl+gl10} consider the effect
of different initialization methods for EM algorithms for mixtures of regressions, including regression
models with random effects.  
\cite{tan+n14} consider variational methods for computation and model choice 
in a generalization of the model of \cite{ng+mwjn06} to allow
covariate dependent mixing weights.  

In the Bayesian nonparametric literature several authors have considered mixed effects
models with Dirichlet process mixture or other nonparametric priors on the distribution of the random effects. 
\cite{kleinman+i98} extend work of \cite{bush+m96} on semiparametric analysis of randomized
block experiments to longitudinal linear mixed effects models.   The focus of their work is on flexible inference
rather than clustering.  \cite{muller+r97} considered nonlinear longitudinal models with Dirichlet
process priors, again focusing on flexible inference rather than clustering.  
\cite{heinzl+t13} consider an EM algorithm for point estimation with
a truncated Dirichlet process prior.  
\cite{deyoreo+rh17} consider a mixture model for datasets in which observations contain both 
ordinal and categorical components.  The variables
are divided into two groups, which they call focus and remainder variables.  Their mixture model allows 
a possibly large number of components for focus variables, and fewer components in modelling
remainder variables.  Although their method is similar to ours in trying to define a focus for the clustering,
their approach focuses on clustering for discrete variables and requires observations
to be vectors of the same dimension for the partitioning.  Our focus is on longitudinal data
where the number of observations and times of observation are not common to all subjects.  
\cite{rigon+hd20} have recently considered a loss-based 
generalized Bayesian approach that can bridge the gap between complex mixture modelling
and loss-based clustering methods while quantifying uncertainty.

As an alternative to mixtures, \cite{booth+ch08} consider a partition model with a multilevel linear mixed model for
observations in each element.  They integrate out the model parameters to obtain a posterior distribution
of the partition which they explore using stochastic search methods.  
Their approach extends an earlier method of \cite{heard+hs06} that does not allow for correlation between
observerations within the same cluster. \cite{delacruz-mesia+qm08} consider quite general mixtures of nonlinear mixed
effects models, similar to earlier work by \cite{pauler+l00}.  The latter authors do not focus on clustering in their work.
%\cite{lenk+d00} is another early contribution to the literature on 
%mixtures of linear mixed models, without particularly focusing on clustering or longitudinal data.  

Parallel to the literature on clustering for longitudinal data, there is closely related work on functional clustering.  
\cite{jacques+p14} give a recent survey.   \cite{james+s03} described one functional data analysis approach
that uses a mixture of linear mixed models.  They considered clustering
using spline basis expansions and note the ease of handling irregularly sampled data using
this approach.  
\cite{shi+w08} consider a finite mixture of Gaussian processes for functional clustering which is useful
when the focus is on response and covariate relationships.  \cite{mcdowell+mvsre18} consider
a Dirichlet process mixture of Gaussian processes, which avoids the need to separately fit models with different
numbers of mixture components. 

The method developed here makes use of linear mixed models, but not through their use as component
models in mixtures or partitions.  
In the next Section, we describe our approach based on mixed predictive replicates and predictive projections.    
In Section 3, we discuss the choice of the number of clusters,
and how we can describe cluster uncertainty using the posterior distribution of the projection.  
Section 4 discusses one synthetic and four real examples with different
features and the performance of our method compared to 
other benchmarks.  Section 5 gives some concluding discussion.

\section{Clustering using mixed models and predictive projections}

Consider correlated data for which the $i$th observation is denoted
$y_i=(y_{i1},\dots, y_{in_i})^\top$, $i=1,\dots, n$.  In this work, usually $y_i$ will be a
response vector for the $i$th individual in a longitudinal study, 
where $y_{ij}$ is a measurement obtained at a time $t_{ij}$, 
$1\leq i\leq n$, $1\leq j\leq n_i$, and times are ordered so that $t_{i1}<\dots <t_{in_i}$.
Let $X_i$ and $Z_i$ be known subject specific design matrices   
(of dimensions $n_i\times p$ and $n_i\times q$ respectively) for fixed and random effects 
respectively for observation $i$.  
Consider a Gaussian linear mixed model of the form
\begin{align*}
	y_i & = X_i\beta+Z_i b_i +\epsilon_i, \;\;\;i=1,\dots, n,
\end{align*}
where $\beta$ denote fixed effect parameters, $b_i$ are random effects, $b_i\sim N(0,G)$ say, 
and $\epsilon_i\sim N(0,\Gamma_i)$.  
We denote by $\eta$ any variance parameters determining $G$ and $\Gamma_i$, so that the parameters
in the model are $\theta=(\beta^\top,\eta^\top)^\top$, and we write $b=(b_1^\top,\dots, b_n^\top)^\top$
for the set of random effects.   In our later examples we will choose
$\Gamma_i=\sigma^2 I$ where $\sigma^2>0$ is a scalar variance parameter
and $I$ denotes the identity matrix.  Although we consider only the case of normally distributed random
effects here, other distributional assumptions are possible, including mixture models, and this is discussed later.  

For Bayesian inference we use a prior $p(\theta)$ on $\theta$. Denote the posterior density of $(\theta,b)$ by
$p(\theta,b|y)$.  We consider a method for clustering based on the use of predictive replicates for 
the original observations.  
%The consideration of predictive replicates with various kinds of replication 
%is common in Bayesian model checking \citep{gelman+ms96}., 
%where they
%are used to judge whether replicated data from the fitted model ``look like'' the observed data.  
%We make a related use of predictive replicates here, where the predictive distributions of replicates are used in
%clustering to judge whether different observations are similar.  
Write $y_i^*$ for the predictive replicate for $y_i$, $i=1,\dots, n$, where 
$y_i^*$ and $y_i$ share the same value of the parameter $\theta$, as well as the same value for a subset of the random effects
$b_{iA}$ say, where we partition $b_i=(b_{iA}^\top,b_{iB}^\top)^\top$.  
The random effect for $y_i^*$
is denoted by $r_i=(b_{iA}^\top,r_{iB}^\top)^\top$, where the part of the random effect $r_{iB}$ for the replicate which is not
shared with $y_i$ is drawn from the conditional prior given $b_{iA},\theta$:  that is, 
\begin{align}
	r_{iB}|b_{iA},\theta & \sim N(G_{AB}^\top G_A^{-1}b_{iA},G_B-G_{AB}^\top G_A^{-1}G_{AB}), \label{reff-rep}
\end{align}
where we have partitioned $G$ according to $b_i=(b_{iA}^\top,b_{iB}^\top)^\top$ as
\begin{eqnarray*}
	G & = \left[\begin{array}{cc} 
		G_A & G_{AB} \\
		G_{AB}^\top & G_B
	\end{array}\right].
\end{eqnarray*}
The purpose of considering these mixed predictive replicates is that the shared random effects will define
relevant variation for forming the clusters in the method we propose,  while 
integrating out the random effects which are not shared
filters out variation considered to be irrelevant.  
Ignoring certain information in order to simplify is an essential part of clustering, and integrating out a subset of random
effects in the mixed predictive distributions makes this explicit in our method.
More precisely, for any $(b_{iA},\theta)$ denote by $p(y_i^*|b_{iA},\theta)$ the predictive distribution 
for $y_i^*$ given $b_{iA},\theta$
after integrating out $r_{iB}$, $i=1,\dots, n$.  Next, consider restricting these predictive distributions to
 a space where there are a finite number $K$ of distinct values for the shared random effects $b_{iA}$, $i=1,\dots, n$.  
This gives a family of approximations to the exact predictive distributions $p(y_i^*|b_{iA},\theta)$.  Within
 our family of approximations, we can find the $K$ distinct values of the shared random effects and an assignment
of these to observations so that our approximate predictive distributions are closest to the 
actual ones in the Kullback-Leibler sense.   Computation of the projection
can be done using a $K$-means type algorithm.   We describe the approach more precisely below.
 
%Clustering of the $y_i$ will be based on the predictive densities of mixed predictive replicates, 
%$p(y_i^*|b_{iA},\theta)$, and we consider Kullback-Leibler projections of these densities onto
%a space where there are only $K$ distinct values among the shared random effects
%$b_{iA}$, $i=1,\dots, N$
%to form a clustering with $K$ clusters.  We discuss further below approximate methods for computation of the projections, 
%as well as development of measures of clustering uncertainty and choice of the number of clusters.

Consider partitioning the columns of $Z_i$ as $Z_i=[Z_{iA},Z_{iB}]$, where $Z_{iA}$ and $Z_{iB}$ are the
columns of $Z_i$ for random effects $b_{iA}$ and $b_{iB}$ respectively.
For the mixed predictive replicates, integrating out $r_{iB}$ gives the conditional density
\begin{align}
	y_i^*|b_{iA},\theta & \sim N\left(X_i\beta+Z_{iA}b_{iA}+Z_{iB}G_{AB}^\top G_A^{-1}b_{iA},
	Z_{iB}\left(G_B-G_{AB}^\top G_A^{-1}G_{AB}\right)Z_{iB}^\top+\Gamma_i\right). \label{predictive}
\end{align}
For a certain value for $(\theta^\top,b_A^\top)^\top$ suppose we 
want to approximate $p(y_i^*|b_{iA},\theta)$ by restricitng to a space where in 
$b_A=({b_{1A}}^\top,\dots,{b_{nA}}^\top)^\top$ there are only $K$ distinct values.  
This gives a clustering of the $n$ subjects into $K$ clusters associated with the
posterior sample $\theta,b_A$.  
Denote the $K$
distinct values among the $b_{iA}$, $i=1,\dots, n$, by $d_{1A}^K,\dots, d_{KA}^K$.  
Write $C_1,\dots, C_K$ for a partition of the set $\{1,\dots, n\}$ into clusters, where $C_j$ contains
the indices of observations in cluster $j$, 
$b_{iA}=d_{jA}^K$ for all $i\in C_j$.  Write $z_i(C)\in \{1,\dots, K\}$ for the value of $j$ such that
$b_{iA}=d_{jA}^K$.    We consider approximating the distribution of predictive replicates
$p(y_i^*|b_{iA},\theta)$ by $p(y_i^*|d_{z_i(C)A}^K,\theta)$, and we want 
to choose $d_{jA}^K$, $j=1,\dots, K$, and $C$ so that this approximation is best in 
the Kullback-Leibler sense.  

The Kullback-Leibler
divergence between distributions with densities $f(y)$ and $g(y)$ is defined when it exists to be
$$\text{KL}(f(y)||g(y))=\int \log \frac{f(y)}{g(y)} f(y)\,dy.$$
We form clusters in our approach by finding Kullback-Leibler projections solving 
the minimization problem
\begin{align}
	& \min_C \min_{d_{1A}^K,\dots,d_{kA}^K} 
	\sum_{i=1}^n  \text{KL}(p(y_i^*|b_{iA},\theta) || p(y_i^*|d_{z_i(C)A}^K,\theta)),  \label{KL-opt}
\end{align}
where $p(y_i^*|b_{iA},\theta)$ is the normal density given at \eqref{predictive}.
Our use of the term ``projection clustering'' in this work
should not be confused with methods in the literature
using this phrase to denote projection of the original data into a lower-dimensional space in a preliminary step.
The Kullback-Leibler divergence considered in (\ref{KL-opt}) is between two multivariate normal distributions 
with a common covariance matrix.  Using the closed-form expression for the Kullback-Leibler divergence
between multivariate normal distributions gives
\begin{align*}
	\text{KL}(p(y_i^*|b_{iA},\theta) || p(y_i^*|d_{z_i(C)A}^K,\theta)) & = 
	\frac{1}{2} (d_{z_i(C)A}^K-b_{iA})^\top 
	Q_i^{-1}
	(d_{z_i(C)A}^K-b_{iA}).
\end{align*}
where
\begin{align*}
	Q_i^{-1} & = \left(Z_{iA}+Z_{iB}G_{AB}^\top G_A^{-1}\right)^\top
	\left\{Z_{iB}\left(G_B-G_{AB}^\top G_A^{-1}G_{AB}\right)Z_{iB}^\top+\Gamma_i\right\}^{-1} 
	\left(Z_{iA}+Z_{iB}G_{AB}^\top G_A^{-1}\right).
\end{align*}

To compute the projection, we use a greedy approach to the optimization where we intialize $C$ and
and then optimize $d_{1A}^K,\dots, d_{KA}^K$ for $C$ fixed, followed by optimization of
$C$ for $d_{1A}^1,\dots, d_{KA}^K$ fixed.  These two steps are iterated until covergence.
This results in a $K$-means type algorithm.
Simple calculus shows that optimization of $d_{1A}^K,\dots, d_{KA}^K$ for fixed $C$ results in 
\begin{align*}
	d_{jA}^K & = \left\{ \sum_{i\in C_j} Q_i^{-1} \right\}^{-1}\left\{\sum_{i\in C_j} Q_i^{-1} b_{iA}\right\}.
\end{align*}
Optimization of $C$ for $d_{1A}^K,\dots, d_{KA}^K$ fixed allocates $i\in C_j$ if
\begin{align*}
	j & = \arg \min_{j'}  (d_{j'A}^K-b_{iA})^\top 
	Q_i^{-1}
	(d_{j'A}^K-b_{iA}).
\end{align*}
We initialize $C$  by choosing $z_i(C)$ uniformly at random from $\{1,\dots, K\}$, for $i=1,\dots, n$.
The clustering algorithm is summarized
as Algorithm 1.  

 \begin{algorithm}[!h] 
\caption{Projection clustering algorithm}
\label{alg:projection clustering}
  \vspace{0.1in} 
  \noindent {\it Inputs:}
  \begin{itemize}
  \item Number of clusters $K$.
  \item Training dataset $y$.
  \item Initial clustering $C^{(0)}$ (obtained by random assignment, for example).
  \item Values for $\theta$, $b_{iA}$, $i=1,\dots, n$ (usually obtained as a draw from their posterior distribution).  
  \end{itemize}
  \noindent{\it Output:}
  \begin{itemize}
  \item Clustering $C^*$.
  \end{itemize}
  \vspace{0.1in}
  
  \noindent{\it Initialization:}
  Set $m=0$, $C=C^{(0)}$.
  
  \noindent{\it Projection clustering:}
  Until a stopping rule is satisfied:
  \begin{enumerate}
  \item Calculate for $j=1,\dots, K$, 
  \begin{align*}
	d_{jA}^K & = \left\{ \sum_{i\in C_j} Q_i^{-1} \right\}^{-1}\left\{\sum_{i\in C_j} Q_i^{-1} b_{iA}\right\}.
\end{align*}
\item For $i=1,\dots, n$, 
allocate $i\in C^*$ if
\begin{align*}
	j & = \arg \min_{j'}  (d_{j'A}^K-b_{iA})^\top 
	Q_i^{-1}
	(d_{j'A}^K-b_{iA}).
\end{align*}
  \item $C=C^*$.  
  \end{enumerate}
\end{algorithm}

\section{Cluster uncertainty and choosing the number of clusters}

The procedure described above produces a clustering based on given values of $(\theta,b_A)$.  In general, we may have 
a set of posterior samples $(\theta^{(s)},b_A^{(s)})$, $s=1,...,S$, in the random effects model.  We can do a clustering
for each posterior draw, and this produces a posterior
distribution on the clustering which describes clustering uncertainty.   For example, we can obtain 
a posterior probability for two  individuals being in the same cluster. 

A difficult question is how to choose the number of clusters.  We consider two approaches.  The first is related to
a method considered for model choice in projection predictive variable selection discussed in \cite{dupuis_variable_2003}.
Let $\text{KL}_K(\theta,b_A)$ denote
the optimized value of the Kullback-Leibler divergence 
$\sum_{i=1}^n \text{KL}(p(y_i^*|b_{iA},\theta)|p(y_i^*|d_{z_i(C)A}^K,\theta))$ for a clustering of size $K$.  Note that 
$\text{KL}_n(\theta,b_A)=0$ and this is the minimum achievable.  Denote by 
$\text{KL}_K$ the average of $\text{KL}_K(\theta,b_A)$ over a set of $S$ posterior samples for $\theta,b_A$.  Then we propose
to choose the number of clusters $K$ as the smallest value of $K$ such that $\text{KL}_K/\text{KL}_1$ is less than 
some small cutoff value $\epsilon$, such as $0.1$.  Since $\text{KL}_K$ 
decreases monotonically in $K$ to its minimum of $0$ at $K=n$, choosing
$K$ in this way chooses the clustering with the fewest clusters such that $\text{KL}_K$ is reduced by $100(1-\epsilon)$\% 
relative to its maximum value.  
This method for choosing the number of clusters requires a choice of $\epsilon$, and an intuitive selection of 
this value relevant to the problem at hand can be difficult.  

The second method investigated for choosing the number of clusters is based on
the notion of clustering stability, using a bootstrap method proposed by \cite{fang_selection_2012}.
Formally, a clustering can be defined as a function $G:Y\rightarrow \{1,\dots, K\}$ where
$Y$ is the space of observations and $K$ is the number of clusters, so a clustering is a function
that maps any observation to a corresponding cluster.  
%Denote an element $\yt\in Y$ 
%by $\yt=(\yt_1,\dots, \yt_\nt)^\top$ 
%where the observations are taken at times 
%$\tt_1,\dots, \tt_{\nt}$.  There are associated design matrices $\Xt$ and $\Zt$ for the fixed
%and random effects for $\yt$ but we will suppress dependence of the clusterings on design matrices 
%in our notation for simplicity.  
Write $G(\cdot;y)$ for a clustering with $K$ clusters obtained from the training data $y$ for $n$ individuals.  
For our mixed model clustering method
the training observations $y_i$, $i=1,\dots, n$, have associated design matrices $X_i$ and $Z_i$, and dependence
of the clustering on these as well as on $n$ and $K$ is suppressed in our notation.  Let $G_1(\cdot;y)$ and $G_2(\cdot;y)$ be two
clusterings with $K$ clusters.  \cite{fang_selection_2012} define the distance between two clusterings
by
\begin{align*}
 d(G_1,G_2) = & P(G_1(\yt;y)=G_1(\yc;y)\;\mbox{and}\;G_2(\yt;y)\neq G_2(\yc;y))+ \\
 & \;\;\;P(G_1(\yt;y)\neq G_1(\yc;y)\;\mbox{and}\;G_2(\yt;y)= G_2(\yc;y)),
\end{align*}
where $\yt, \yc$ are observations drawn independently from the same population as the training samples.  
So the distance between clusterings is defined as the probability that two independent draws 
from the population
will be clustered differently by the two methods.

Now consider the case where the clusterings $G_1$ and $G_2$ are obtained by the same clustering
algorithm, but using different training data.  
Following \cite{fang_selection_2012} define the clustering instability to be 
$$E(d(G_1(\cdot;y'),G_2(\cdot;y''))),$$
where the expectation is with respect to the distribution of two independent training samples of size $n$ from 
the population, denoted here by $y'$ and $y''$.  \cite{fang_selection_2012} 
approximate the expectation by drawing two independent
bootstrap samples of size $n$ from the original training sample, 
computing the proportion of the original training sample pairs for which
the clusterings for the bootstrap samples disagree, and then averaging these over $B$ bootstrap replicates.  
For the choice of $B$, \cite{fang_selection_2012} suggest that $B=20$ or $50$ can be adequate 
in their experience, and later we use $B=100$ in our examples.

In our clustering method, suppose we first represent each observation through
its fitted mean for the predictive replicates at the union of times for all subjects.  
Denote the estimated instability for $K$-means clustering applied to these fitted means
for bootstrap sample $y_b$ by $I_K(y_b)$, $b=1,\dots, B$, and denote by $\text{I}_K$ the average
of these measures over the $B$ bootstrap samples.  Clustering based on fitted means for replicates
is used to reduce the computational burden that would result from the need to average over both posterior
samples and bootstrap replicates in a more direct application of the method
of \cite{fang_selection_2012} here.
We adapt the method of 
\cite{fang_selection_2012} to choose the number of clusters as
$$K=\arg\min_{2\leq k\leq K_{\text{max}}} \text{I}_K$$
where in the minimization $K_{\text{max}}$ is the maximum allowable cluster
size and the choice $K=1$ is excluded since in the trivial case of one cluster there is
no instability.  

In our later examples we modify the bootstrap approach to achieve greater parsimony in the number
of clusters by choosing the number of clusters as small as possible subject to the instability
being no less than half its maximum value.  We use $K_{\text{max}}=30$ and choose 
$$K=\min \left\{k:  \text{I}_k\geq 1/2 \max_{2\leq l\leq 30} \text{I}_l \right\}.$$
Without this adjustment the clustering instability does not reach a minimum value for $K$ less than $30$ in our
examples, and it is often hard to interpret such a large number of clusters. 

\section{Examples}

%evaluation of methods: examples: summary of description; subsection-dataset and results
We demonstrate performance of our method for five examples with different features.  
The first example uses a synthetic dataset to illustrate how our method can reveal
structure at different scales through the choice of the shared random effects in
constructing predictive replicates.  The remaining examples involve real data.  
Examples two and three have only a small number of observations per subject, and no additional covariates apart from time.  
The fourth example has a large number of observations per subject, and we reduce dimension by transforming
each sequence to a power spectrum at 40 different frequencies. 
The fifth example includes additional covariates as fixed effects in the model.  
All of the real examples have the same number of observations per subject to meet
the requirements of the competing benchmark methods we consider.  However, implementation with unbalanced
data is demonstrated for our method in Example 5, where gaps are randomly introduced.  
Data and analysis code for examples are available at \url{https://github.com/maoyinan/Projection-Clustering}. 

\subsection{Datasets}

We give some background on the five examples first, before discussing the clustering results.

\paragraph{Example 1: Synthetic dataset}

Our first example shows that the choice of random effects which are shared with the replicates
in our method can allow the user to focus on features of interest for clustering.
We consider data generated in four groups.  The mean for each subject is a sum of two cosine basis functions
with random frequencies for each individual.  For each individual, one basis function is a ``low frequency'' term
and one basis function is a ``high frequency'' term.  The coefficient for each basis function 
can be large or small in magnitude, so that the low or high frequency signal can be strong or weak.  
The coefficients vary according to the four groups.  The four groups are strong low and strong
high frequency (SLSH), strong low and weak high frequency (SLWH), weak low and strong high frequency
(WLSH) and weak low and weak high frequency (WLWH).  

We construct later two clustering methods in our projection framework.  One distinguishes between strong and weak low
frequency behaviour, while ignoring the high frequencies.  The other distinguishes between strong and weak high
frequency behaviour, while ignoring the low frequencies.  The goal here is not to ``correctly'' find four classes, but
rather to focus only on a certain type of variability in forming clusters (in this case strong/weak low frequency
or strong/weak high frequency signal).
The data for this example are generated in the following way.  
For each subject, there are $T=40$ observations at times $t=1/T,2/T,\dots, (T-1)/T,1$.  Then for subject $i$
the responses are generated as
\begin{align}
 y_{it} & = \beta^{(1)}_{\delta_i} \cos(\pi w^{(1)}_i t) +  \beta^{(2)}_{\delta_i} \cos(\pi w^{(2)}_i t) + \epsilon_{it} , \label{mean-model}
\end{align}
where $w^{(1)}_i$ and $w^{(2)}$ are discrete uniform on $\{1,2,3\}$ and $\{7,8,9\}$ respectively, 
$\epsilon_{it}\sim N(0,0.1)$, $\delta_i=j$ if individual $i$ is in 
group $j\in \{1,2,3,4\}$, where groups $1,2,3$ and $4$ are the SLSH, SLWH, WLSH and WLWH groups respectively, 
and 
$$(\beta^{(1)}_j, \beta^{(2)}_j) = 
\begin{cases} 
	(1,1) &\mbox{if } j= 1 \\ 
	(1,0.1) & \mbox{if } j =2\\
	(0.1,1) & \mbox{if } j =3\\
	(0.1,0.1) & \mbox{if } j =4
 \end{cases}  .$$
On the right-hand side of (\ref{mean-model}), the first and second terms are low and high frequency signals. 
The data are plotted in Figure \ref{fig:series1} in the Appendix.

\paragraph{Example 2: Crop image}

This example comprises crop image data 
obtained from the UCR Time Series Classification Archive \citep{dau_ucr_2019}.
Each observation is a time series associated with a pixel from a satellite image, where the images
at different times are corrected so that a given pixel corresponds to the same spatial region in all 
images.  The time series are
of length 46, and show the temporal evolution.  Class labels are known specifying
the land usage.  In the full dataset there are 24 true classes.  Here we sample
5 of the classes randomly and use the first 30 observed series within each class, giving 150 time series of 
length 46 in total.  The data are shown in Figure \ref{fig:series1} in the Appendix.
%Set A is chosen to be Fourier basis of order $1-3$, $4-10$, $11-30$ as low frequency level, medium frequency level, and high %frequency level respectively.
 
\paragraph{Example 3: DNA synchrony of yeast cells}
This example considers gene expression data where each time series gives gene expression level 
over time relative to a control sample in yeast cells of 5 stages \citep{spellman_comprehensive_1998}.  
Each series contains 18 records measured 7 minutes apart.  We consider 30 genes in each cell stage, 
giving 150 time series of length 18, with the true class given by the 5 stages.  

%Fourier basis of order $1$ to $18$ were evenly distributed into the $3$ frequency levels as set A for projection clustering.

\paragraph{Example 4: EEG signals during sleep}
This example concerns electroencephalogram (EEG) recordings during sleep for different sleep stages 
(wake, S1, S2, S3, S4, REM, body movements). Records in channels Fp3-F4 of a bruxism patient (brux2) were downloaded from 
the CAP sleep database archived on PhysioNet \citep{terzano_atlas_2001}. Raw EEG recordings were sampled at $512$Hz, from which we randomly sampled $30$ second segments. Due to the high noise level of EEG signals, they were further mapped into frequency spectra below $40$Hz via a fast Fourier transform. The final data consists of 
$89$ frequency spectra at 40 different frequencies.  

%We used a B-spline basis of order $6$, $12$, $18$ as set A for low frequency level, medium frequency level, and high frequency %level respectively, instead of fourier basis for this particular example, as frequency spectrum does not contain periodical patterns.

\paragraph{Example 5: Activity recognition from accelerometer data}
This example, from the UCI machine learning repository \citep{DuaGra2019}, concerns an activity dataset of  $15$ subjects performing  $7$ activities including  1: Working at Computer , 2: Standing Up, Walking and Going Up/Down stairs , 3: Standing, 4: Walking , 5: Going Up/Down Stairs ,  6: Walking and Talking with Someone, 7: Talking while Standing \citep{casale_personalization_2012}. 
A single chest-mounted accelerometer recorded acceleration data in $3$ dimensions and measured at $52$Hz. Data from all participants in the vertical dimension was pooled together for the activity recognition task, 
where segments containing more than $50$ points were truncated at $50$, and segments with less than $10$ data points were discarded. 
%The $3$ frequency levels correspond to fourier basis of $1-3$, $4-10$, and $11-30$ as random effects.
In this example, we considered two other versions of the data to demonstrate the ability of our method to handle covariates
and missing or unbalanced data. In the first variant, acceleration data in the two other dimensions were included as fixed effects and modeled along with random effects. In the second variant, we randomly introduced $10\%$ missingness into the original version of the data.

\vspace{0.1in}
For all examples, times were scaled to lie in the range $[0,1]$, and the responses were scaled to have mean
zero and variance one.  
Linear mixed models were fitted using MCMC using the R package \texttt{rstan}  \citep{team_rstan_2021}, with default prior
settings.  We ran $4$ chains for $2000$ iterations with $1000$ burn in, obtaining $4000$ MCMC samples 
in each case with no thinning. All Fourier basis terms are included as random effects.
Denote the $j$th row of the design matrices $X_i$ and $Z_i$ in the mixed model by $X_{ij}$ and $Z_{ij}$. 
The $j$th rows of $Z_{i,A}$ and $Z_{i,B}$ are denoted $Z_{ij,A}$ and $Z_{ij,B}$ respectively. 
We specify $X_{ij}=[1]$, while $Z_{ij}$ and $Z_{ij,A}$ are example specific and discussed below in each case.  
In all our examples, there is a ``true'' class label available, and we make use of these in evaluating
the clustering methods we consider.  However, in most cases in practice
there are no true class labels, and even if there are such labels recovering them may not be the purpose
of a cluster analysis \citep{akhanli_comparing_2020}.  
As we have emphasized, a main advantage of our method is the ability to specify 
what aspects of the data define clusters through the choice of random effects used in defining 
mixed predictive replicates.  We demonstrate this first, using the synthetic data example 1.  

\subsection{Synthetic example results}

Write $F(j,t)=\cos(\pi j t)$, and let $Z_{ij}=[F(0,t_{ij}), F(1,t_{ij}),\dots, F(9,t_{ij})]$, where $t_{ij}=t_j=j/T$, $j=1,\dots, T$,
are the observation times for subject $i$, with $T=40$.  Note that $F(0,t_{ij})=1$ is an intercept term.  
With this choice of $Z_{ij}$, the covariates appearing as random effects
are cosine basis terms with different frequencies.  
We consider applying our clustering method with $X_{ij}=[1]$ 
and four different choices of $Z_{ij,A}$;  $Z_{ij,A}=Z_{ij}$ (all frequencies), 
$Z_{ij,A}=[F(0,t_{ij}), F(1,t_{ij}), \dots, F(3,t_{ij})]$ (low frequencies),
$Z_{ij,A}=[F(4,t_{ij}), \dots, F(6,t_{ij})]$ (intermediate frequencies),
and $Z_{ij,A}=[F(7,t_{ij}), \dots, F(9,t_{ij})]$ (high frequencies). 
Recall that the generative process for this example has four groups, 
strong low and strong
high frequency (SLSH), strong low and weak high frequency (SLWH), weak low and strong high frequency
(WLSH) and weak low and weak high frequency (WLWH).  We fix the number of clusters to $4$.
Choosing the number of clusters from the data is considered later, but here
we illustrate the properties of our clustering approach in a simple setting.

When $Z_{ij,A}=Z_{ij}$, we should do well in distinguishing all four groups.  If we 
cluster with low frequency basis terms in $Z_{ij,A}$, we should distinguish well between
groups with different low frequency behaviour, but not high frequency behaviour.  
If we cluster with high frequency terms in $Z_{ij,A}$, we should distinguish well between
groups with different high frequency behaviour, but not low frequency behaviour.
Finally, with intermediate frequencies, we exclude the important information for distinguishing
between all the groups, and might not expect the method to distinguish with confidence between any of
the groups.

\begin{figure}[H]
	\begin{center}
		\includegraphics[width=140mm]{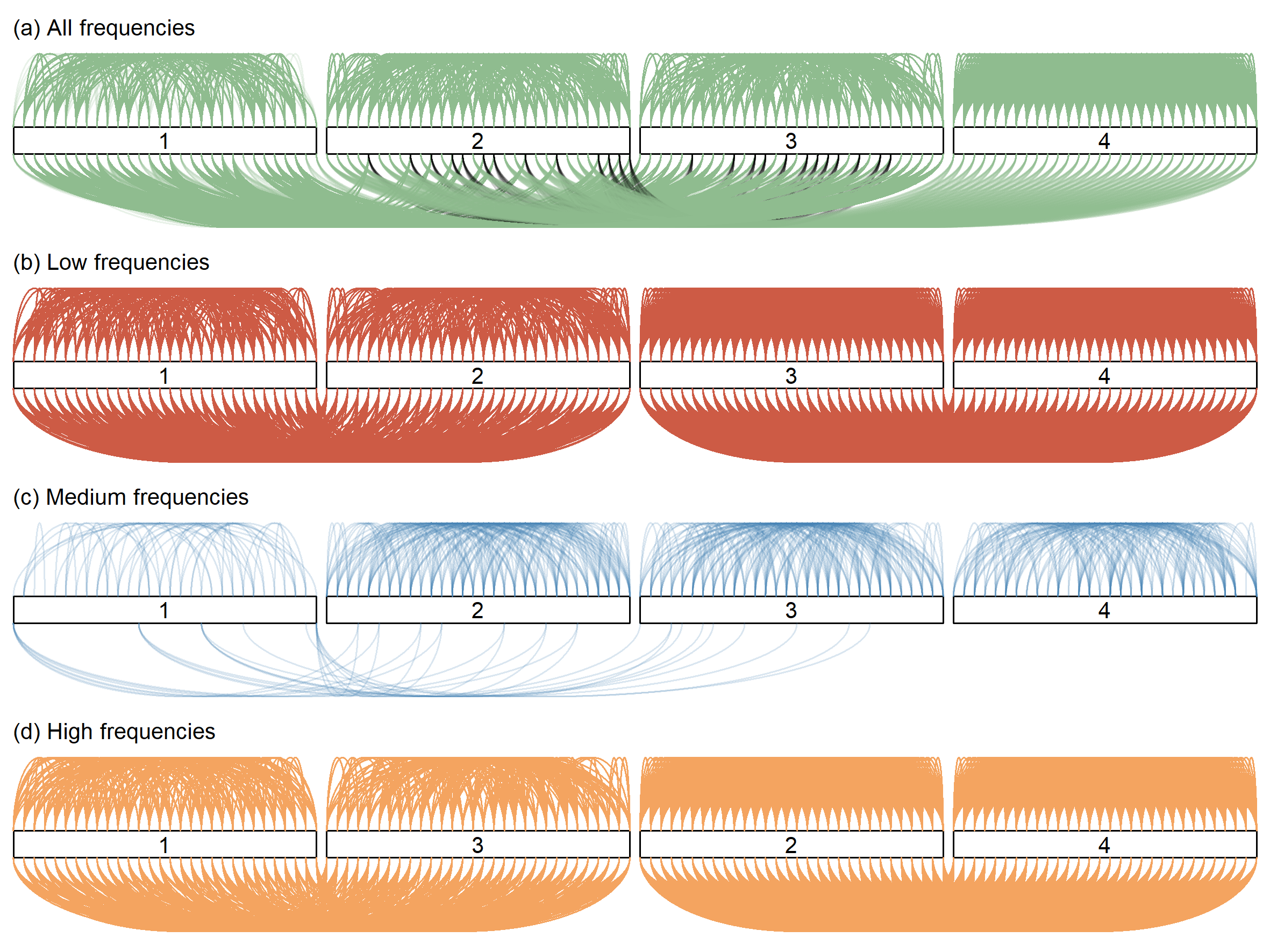}
	\end{center}
	\caption{Pairwise coincidence probabilities for (a) all frequency, (b) low frequency, (c) intermediate frequency and (d) high frequency cases for $Z_{i,A}$ for synthetic example (class labels reordered for clarity).  Interpretation is discussed
	further in the text.}
	\label{fig:clusterProb1}
\end{figure}

We summarize the results by pairwise coincidence probabilities.  The pairwise coincidence probability for subjects $i$ and $j$
is the probability that they are clustered together.    
The probabilities are estimated based on 4,000 MCMC samples.  
Figure \ref{fig:clusterProb1} summarizes the results.  Transparent curves connecting two subjects indicates weak coincidence probability between them in the range $0.5$ and $0.8$, and a solid curve indicates a probability $>0.8$. The subjects are arranged together if they belong to the same group for easier visualization. Subjects correctly clustered in the same group are linked by curves above the group label.  Pairs which are clustered wrongly together but from two similar groups 
are shown as colored lines below the group labels. Similar groups are ones where the low frequency behaviour is the same, or 
the high frequency behaviour is the same.  
Pairs clustered wrongly together with probability $>0.8$ from groups that are not similar are shown as black.  

We make a number of observations.  First, for the low frequency choice of $Z_{ij,A}$, we never
wrongly cluster a strong signal low frequency observation with a weak low frequency observation with high
confidence (i.e. no black links
below the labels).  Similarly, for the high frequency choice of $Z_{ij,A}$, 
Figure 1 (d) shows there are no cases of a strong high frequency observation 
wrongly clustered with weak high frequency observation
(again, no black links below the labels).  In Figure 1 (d) the classes are ordered differently in the plot for clarity.  
For the intermediate frequency case, there are few large pairwise coincidence probabilities at all, showing
the loss of information about the true groups 
that occurs when we exclude both high and low frequency basis terms from $Z_{ij,A}$.
When we use all frequencies, the black links below the graph show some cases of individuals
clustered together from groups which are not similar.
This shows that the filtering of the noise done by integrating out some of the random effects helps
for the targeted goal of distinguishing strong/weak low frequency or strong/weak high frequency.
This example shows how the choice of $Z_{ij,A}$ can allow the analyst to successfully focus the clustering method 
on features of the variability of interest for forming clusters.

\subsection{Real Example results}

\begin{figure}[H]
	\begin{center}
		\includegraphics[width=100mm]{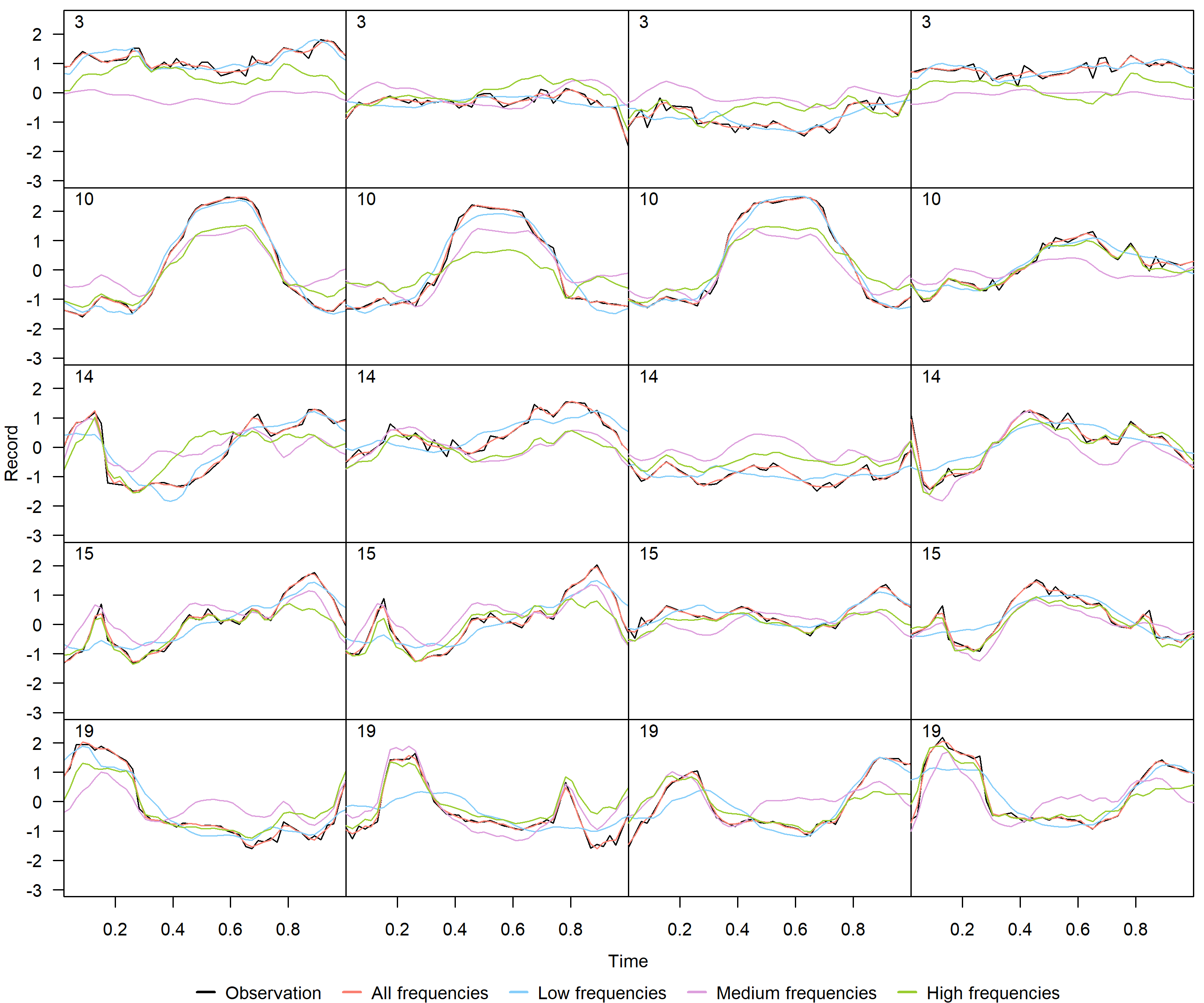}
	\end{center}
	\caption{\label{fig:fitted2}Fitted means for mixed predictive replicates for crop data and low frequency, intermediate frequency, high frequency and all frequency cases for $Z_{i,A}$.  The number in the top left corner of each graph is the class label.   Four randomly chosen observations for each class are chosen for plotting.}
%\end{figure}

%\begin{figure}[H]
	\begin{center}
		\includegraphics[width=100mm]{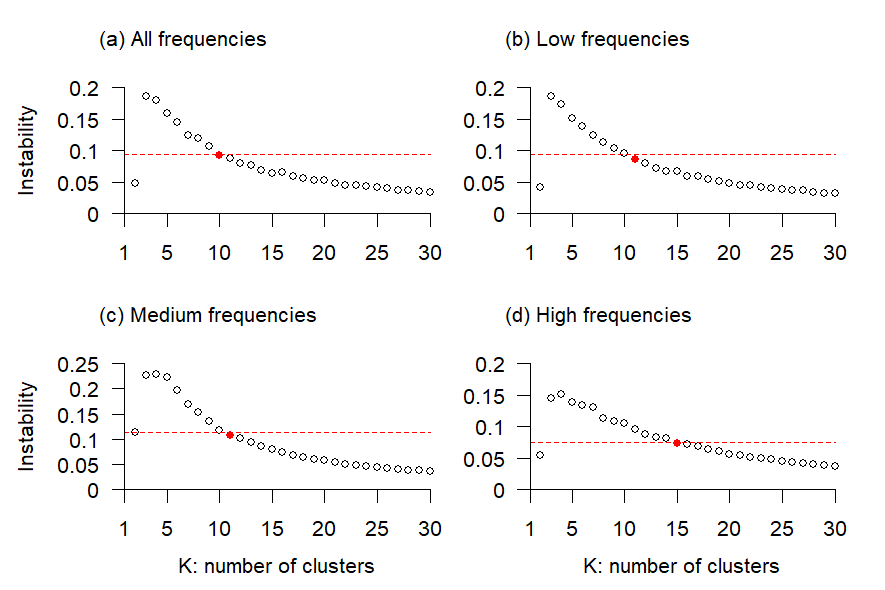}
	\end{center}
	\caption{\label{fig:bootstrap2}Plot of instability $\text{I}_K$ versus $K$ for (a) all frequency, (b) low frequency, (c) intermediate frequency
	and (d) high frequency cases for $Z_{i,A}$ for crop example.}
\end{figure}

We consider the four real examples next, starting with Example 2. 
Here $X_{ij}=[1]$ and $Z_{ij}=[F(0,t_{ij}), F(1,t_{ij}), \dots, F(30,t_{ij})]$.  After fitting the mixed model, 
Figure \ref{fig:fitted2} shows
the fitted means for the mixed predictive replicates in four cases:  1) $Z_{ij,A}=Z_{ij}$, 2) $Z_{ij,A}=[F(0,t_{ij}), F(1,t_{ij}), \dots, F(3,t_{ij})]$ (low frequency case), 3) $Z_{ij,A}=[F(4,t_{ij}), \dots,F(10,t_{ij})]$ 
(intermediate frequency case) and 4) $Z_{ij,A}=[F(11,t_{ij}),\dots, F(30,t_{ij})]$ (high frequency case).  
The number plotted in the top left
of each figure is the true class label; there are five different values for this label since we randomly sampled
5 of the classes from the original dataset.  
Four randomly chosen observations for each class are chosen for plotting.

For each choice of $Z_{ij,A}$, Figure \ref{fig:bootstrap2} shows a plot of $\text{I}_K$ for the bootstrap method
against $K$, and the choice of $K$ as small as possible
subject to $\text{I}_K$ being no less than 50\% of its maximum value.  
This results in $10$ clusters chosen for the case of all frequencies for $Z_{i,A}$, 
$11$ clusters for the low frequency terms for $Z_{ij,A}$, 
$11$ clusters for the intermedicate frequency for $Z_{ij,A}$ and $15$ clusters for high frequency 
terms for $Z_{ij,A}$.  
Using the KL-divergence loss method of choosing the number of clusters described in Section 3 with $\epsilon=0.1$,  
gives a large number of clusters for each case (Figure \ref{fig:KL2} in the Appendix).

\begin{figure}[H]
	\begin{center}
		\includegraphics[width=130mm]{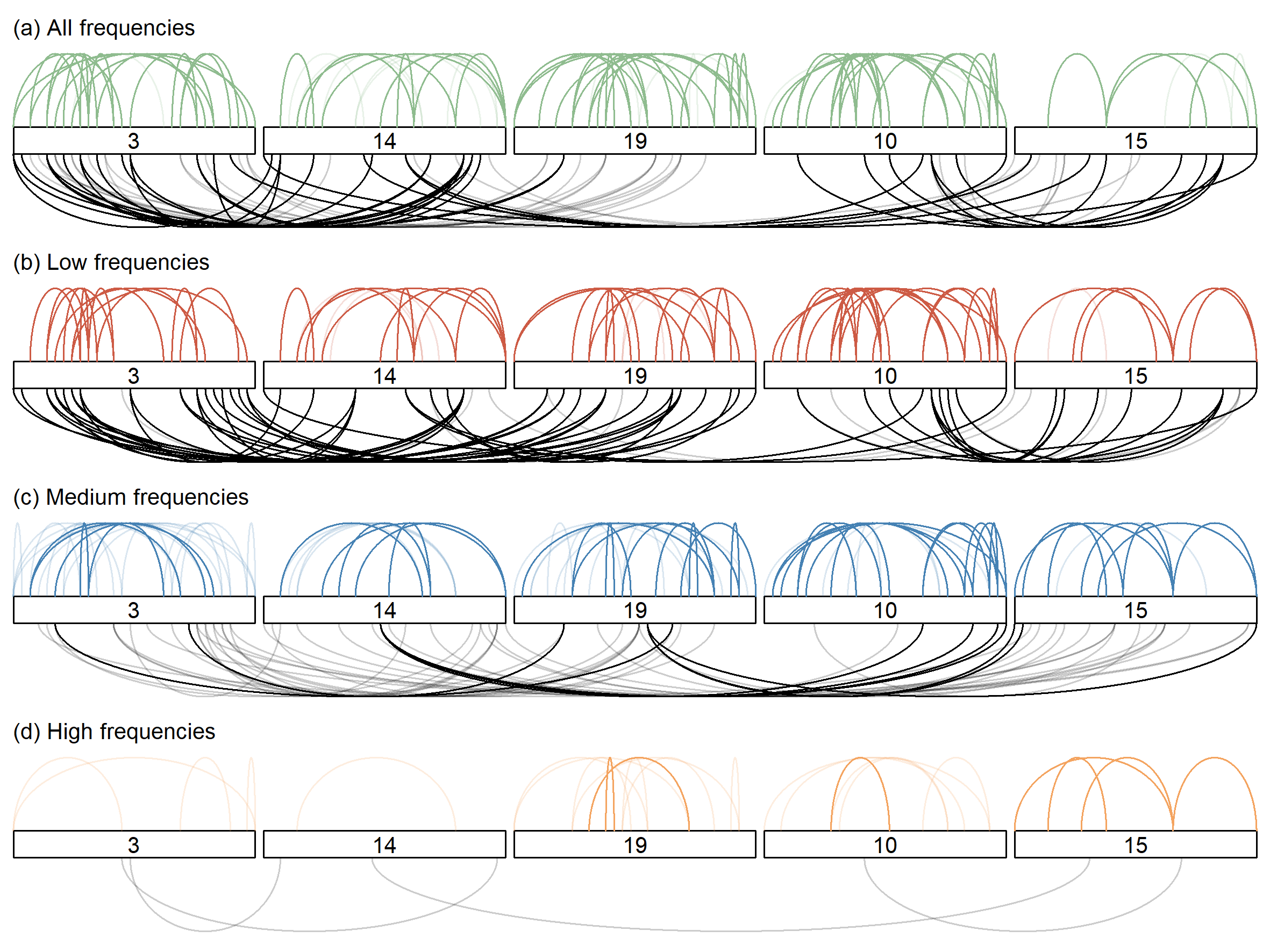}
	\end{center}
	\caption{\label{fig:clusterProb2}Pairwise coincidence probabilities for (a) all frequency, (b) low frequency, (c) intermediate frequency and (d) high frequency cases for $Z_{i,A}$ for crop example.  Interpretation is discussed
	further in the text.  Only 10\% of the links are plotted, randomly chosen.}
\end{figure}

Pairwise coincidence probabilities for our clustering method 
are shown in Figure \ref{fig:clusterProb2}, where the interpretation
of this plot is similar to before for the synthetic data.  For clarity we plot only a randomly chosen 10\% of the
links in the graph.  
Subjects correctly clustered in the same group are linked by curves above the group label, while those wrongly clustered in different groups are drawn below, with black links for wrong classifications for classes that are not similar. 
We make the following observations.  
First, Figure \ref{fig:clusterProb2} can tell us how informative variation at different scales is for
distinguishing between the two classes.  The low and intermediate frequency cases for $Z_{ij,A}$ 
result in a slightly better clustering than the high frequency case, and this can be confirmed quantitatively in the 
next section where we compare our methods with other benchmarks using the Rand index and adjusted Rand index.
Second, we can see that for the low frequency choice of $Z_{ij,A}$ classes 3, 14 and 19 are hard to distinguish, but
for the intermediate frequency case they are more easily distinguished (fewer black links below the labels for these classes).

Similar pairwise clustering results for Examples 3-5 are included in the Appendix, and for these cases 
we use the bootstrap method described in Section 3 for choosing the number of clusters.  
Plots showing the cluster choice via the bootstrap method for different cases are shown 
in Figures \ref{fig:bootstrap3}-\ref{fig:bootstrap5} in the Appendix.  
The bootstrap method tends to give a smaller number of clusters 
than the method based on the loss in KL divergence due to projection, as shown in  
Figure \ref{fig:KL2} in the Appendix for the crop data.

For Example 3,  we chose  
$Z_{ij}=[F(0,t_{ij}), F(1,t_{ij}), \dots, F(18,t_{ij})]$, and low, intermediate and high frequency
choices of $Z_{ij,A}$ are $Z_{ij,A}=[F(0,t_{ij}), F(1,t_{ij}),\dots,F(6,t_{ij})]$, $Z_{ij,A}=[F(7,t_{ij}), \dots, F(12,t_{ij})]$ and $Z_{ij,A}=[F(13,t_{ij}),\dots, F(18,t_{ij})]$ respectively.  
For this example, most of the pairs with high pairwise coincidence probability are clustered in the same 
group or similar groups (next or previous cell stage). Only a few curves appear to wrongly cluster subjects into different groups. Among all the choices for $Z_{ij,A}$, the 
low frequency case (Figure \ref{fig:clusterProb3} (b) in the Appendix) results in better clustering in terms of the true class labels.
Again, this can be confirmed in the comparisons with other benchmarks in the next Section.  This is an example
where filtering out the noise by integrating out some of the random effects actually results in a more accurate clustering
in terms of the original class labels.  

For Example 4, observations contain $30$ seconds of EEG data sampled at $512$Hz, so that there
are $J=30\times 512$ observations per subject.  We apply a discrete Fourier transform
to reduce dimension to a set of 40 frequencies by taking the power spectrum for 
frequencies below $40$Hz: 
$\text{PS}(\omega_k)= |1/2h \sum^{\omega_k+h}_{\omega_k-h} \text{DFT}(\omega_k)|^2$, $k=1,\dots,40$, where 
for a signal $Y(j)$, $j=1,\dots, J$, 
$\text{DFT}(\omega_k)=\sum^{J}_{j=1}  Y(j) \exp{(-2\pi i(j-1)(k-1)/J)}$ where $h=0.5$ and $\omega_k=(k-1)/40$.
Our observation $y_i$ don't correspond to observations over time in this example, 
but are frequency spectrum values at different frequencies
$\omega_k$, $k=1,\dots, 40$. 

Here we consider a B-spline basis for fitting the power spectra.  Let $B(\omega_{ij})$ be a row vector of $30$ cubic B-spline
basis functions obtained using equally spaced knots evaluated at the $j$th frequency for observation $i$.  
The basis functions are ordered according to the knot points, so if we write $B_k(\omega_{ij})$ for the $k$th
entry of $B(\omega_{ij})$, basis functions with lower index are locally fitting lower frequencies.  
We write $B(k\colon l,\omega_{ij})$ for the row vector obtained from $B(\omega_{ij})$ by extracting components $k$ to $l$
inclusive.  Let $Z_{ij}=[1,B(\omega_{ij})]$, and to capture low, intermediate and
high frequency behaviour
in our clustering method, we set $Z_{ij,A}=[1,B(1\colon 5,\omega_{ij})]$, $Z_{ij,A}=[B(6\colon 15,\omega_{ij})]$ and
$Z_{ij,A}=[B(16\colon 30,\omega_{ij})]$ respectively.  
Figure \ref{fig:clusterProb4} in the Appendix shows that the intermediate frequency choice for $Z_{ij,A}$ 
is best for clustering groups 0 and 5. 
%In choosing cluster numbers, the KL threshold method tends to pick a high value when the KL curve levels off slowly. 

For Example 5, we define $Z_{ij}=[F(0,t_{ij}), F(1,t_{ij}), \dots, F(30,t_{ij})]$, and our 
low, intermediate and high frequency choices for $Z_{ij,A}$ are
$Z_{ij,A}=[F(0,t_{ij}), F(1,t_{ij}), \dots, F(5,t_{ij})]$, $Z_{ij,A}=[F(6,t_{ij}), \dots, F(10,t_{ij})]$ and $Z_{ij,A}=[F(11,t_{ij}), \dots, F(30,t_{ij})]$ respectively.  
In Example 5, the third activity type is most successfully identified with little noise in the low frequency case (see Figure \ref{fig:clusterProb5} in the Appendix).

\subsection{Comparison with other benchmarks}

Now that we have examined how our method can reveal structure at different scales through the choice
of $Z_{i,A}$, we examine clustering accuracy quantitatively in terms of the true class labels by using the
Rand Index \citep{rand71} and adjusted Rand Index \citep{hubert+a85} and compare our method to 
some other benchmarks. 
For our method, the Rand and adjusted Rand index values reported are the average values
obtained over $4000$ clusterings obtained from different MCMC samples.  In each case, we
compare our method with the following alternatives: 
(1) HC\_dist: 
hierarchical clustering based on an integrated periodogram-based method as dissimilarity measure \citep{montero_tsclust_2014}
(2) HC\_pred: 
 hierarchical clustering based on a prediction density-based method as dissimilarity measure \citep{montero_tsclust_2014} 
(3) BHC: 
Bayesian model-based hierarchical clustering with accounting for uncertainty using the Dirichlet process \citep{savage_rbhc_2009} 
(4) KML: 
$K$-means for longitudinal data \citep{genolini_kml_2011}
(5) Mclust: 
finite Gaussian mixture model under Bayesian framework estimated by Estimation-Maximisation \citep{scrucca_mclust_2016}
(6) VC: 
clustering based on Bayesian mixtures of linear mixed models estimated via variational inference \citep{tan+n14}.
The first five benchmarks can be applied with corresponding R packages but they all require equal time sampled data. The last benchmark is flexible about input data and has R code available online. 

For our projection clustering method, the case of all frequencies, low frequencies, intermediate frequencies
and high frequencies for $Z_{i,A}$ are denoted as PC1, PC2, PC3 and PC4 respectively.  
Table \ref{tab:rand1} compares our method with the other benhcmark methods for Examples 1-3, and
Table \ref{tab:rand2} compares PC1, PC2, PC3 and PC4 against each other and VC for Example 5, 
where additional covariates and missingness have been added.  The other benchmark methods are not applicable
in these cases.  In Table \ref{tab:rand1}, the methods PC1 and PC2 are competitive with the best
benchmark methods. In Table \ref{tab:rand2}, the addition of covariates allows a small increase in accuracy, 
while the introduction of missingness causes little deterioration in accuracy of the clustering for capturing
the true class labels. 

\begin{table}

\caption{\label{tab:rand1}Rand and adjusted Rand indices for different clustering methods for examples 2-4. The methods compared are described in the text.}
\centering
\begin{tabular}[t]{l|r|r|r|r|r|r|r|r|r|r}
\hline
Example & HC\_dist & HC\_pred & BHC & KML & Mclust & VC & PC1 & PC2 & PC3 & PC4\\
\hline
\multicolumn{11}{l}{\textbf{Rand Index}}\\
\hline
\hspace{1em}Eg2 & 0.57 & 0.67 & 0.78 & 0.81 & 0.80 & 0.59 & 0.82 & 0.80 & 0.77 & 0.76\\
\hline
\hspace{1em}Eg3 & 0.64 & 0.66 & 0.69 & 0.65 & 0.64 & 0.70 & 0.68 & 0.69 & 0.66 & 0.66\\
\hline
\hspace{1em}Eg4 & 0.75 & 0.70 & 0.74 & 0.78 & 0.77 & 0.68 & 0.79 & 0.78 & 0.80 & 0.75\\
\hline
\multicolumn{11}{l}{\textbf{Adjusted Rand Index}}\\
\hline
\hspace{1em}Eg2 & 0.10 & 0.04 & 0.35 & 0.41 & 0.40 & 0.10 & 0.45 & 0.38 & 0.30 & 0.29\\
\hline
\hspace{1em}Eg3 & 0.02 & 0.05 & 0.09 & 0.13 & 0.11 & 0.25 & 0.17 & 0.12 & 0.06 & 0.03\\
\hline
\hspace{1em}Eg4 & 0.31 & 0.13 & 0.24 & 0.34 & 0.34 & 0.04 & 0.34 & 0.31 & 0.36 & 0.23\\
\hline
\end{tabular}
\end{table}

\begin{table}

\caption{\label{tab:rand2}Rand and adjusted Rand indices of different clustering methods for Example 5.  
	The methods compared are described in the text.  Eg5 is the case of the original data, Eg5M introduces 
	additional fixed effects in the model (accelerometer data in two other directions) 
	and Eg5G treats 10\% of the original observations as missing}
\centering
\begin{tabular}[t]{l|r|r|r|r|r}
\hline
Example & VC & PC1 & PC2 & PC3 & PC4\\
\hline
\multicolumn{6}{l}{\textbf{Rand Index}}\\
\hline
\hspace{1em}Eg5 & 0.68 & 0.66 & 0.68 & 0.46 & 0.36\\
\hline
\hspace{1em}Eg5M & 0.68 & 0.65 & 0.68 & 0.57 & 0.46\\
\hline
\hspace{1em}Eg5G & 0.67 & 0.64 & 0.67 & 0.46 & 0.41\\
\hline
\multicolumn{6}{l}{\textbf{Adjusted Rand Index}}\\
\hline
\hspace{1em}Eg5 & 0.04 & 0.05 & 0.04 & 0.06 & 0.05\\
\hline
\hspace{1em}Eg5M & 0.04 & 0.06 & 0.06 & 0.02 & 0.04\\
\hline
\hspace{1em}Eg5G & 0.05 & 0.04 & 0.06 & 0.06 & 0.04\\
\hline
\end{tabular}
\end{table}

\section{Discussion}

We have developed a new model-based clustering method based on mixed predictive replicates and predictive
projections.  The method fits a linear mixed model, and then defines predictive replicates for each observation
where a subset of random effects is shared with the original observations with the other random effects
drawn from the conditional prior.  Considering predictive projections for the mixed predictive distributions
of the replicates, we project onto a space where the number of distinct values for the shared random
effects is finite, defining different clusterings.  The main strength of the method is the way it gives the analyst
flexibility to define what information should be used in defining the clustering, through the choice of shared random
effects for defining replicates.

There are several ways this work could be extended.  We restricted here to fitting a linear mixed model with
Gaussian random effects, but non-Gaussian distributions for the random effects are easily considered.  
Distributions for the random effects such as finite Gaussian mixtures, multivariate $t$ or skew normal 
having a conditionally Gaussian formulation are easy to use with
our method where the latent variables in the conditional Gaussian representation can
be generated by MCMC.   It would also be possible to consider clustering for discrete data based on
generalized linear mixed models, although the computation of projections is more difficult in this case.  
The methods described in \cite{catalina_projection_2020} for projection predictive model selection in 
generalized linear and additive mixed models could possibly be used here.

%\section*{Acknowledgements}
%Data and analysis code for examples are available at \url{https://github.com/maoyinan/Projection-Clustering}. 
\section*{Disclosure Statement} 
All authors declare no financial conflict. 

\bibliographystyle{chicago}
\bibliography{Reference_DN,Reference2}

\newpage

\section*{Appendix - Additional Figures}

\begin{figure}[H]
	\begin{center}
		\includegraphics[width=110mm]{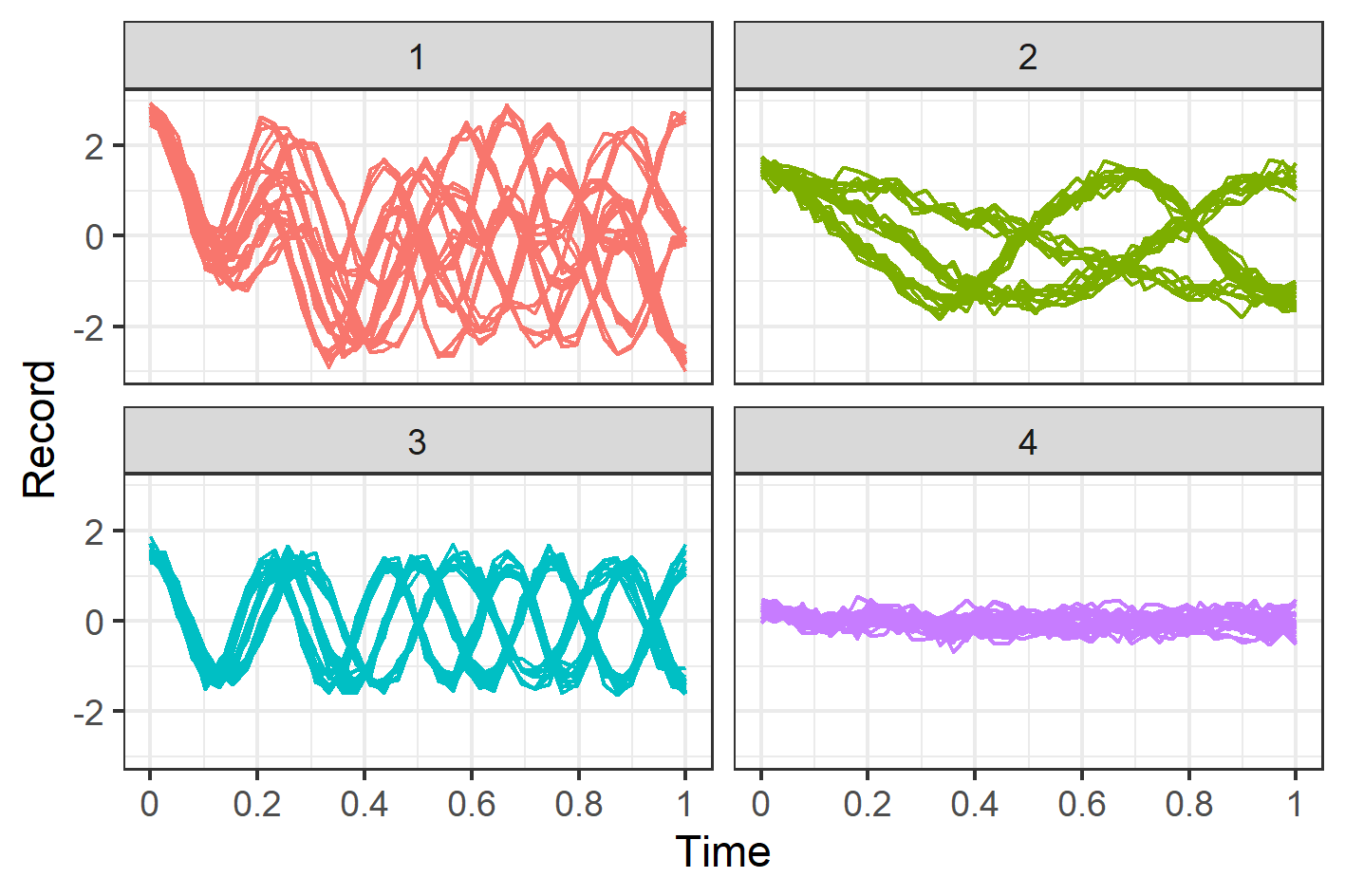}
	\end{center}
	\caption{\label{fig:series1}Time series plots of observations within the four different groups for the synthetic data. }
%\end{figure}

%\begin{figure}[h]
	\begin{center}
		\includegraphics[width=110mm]{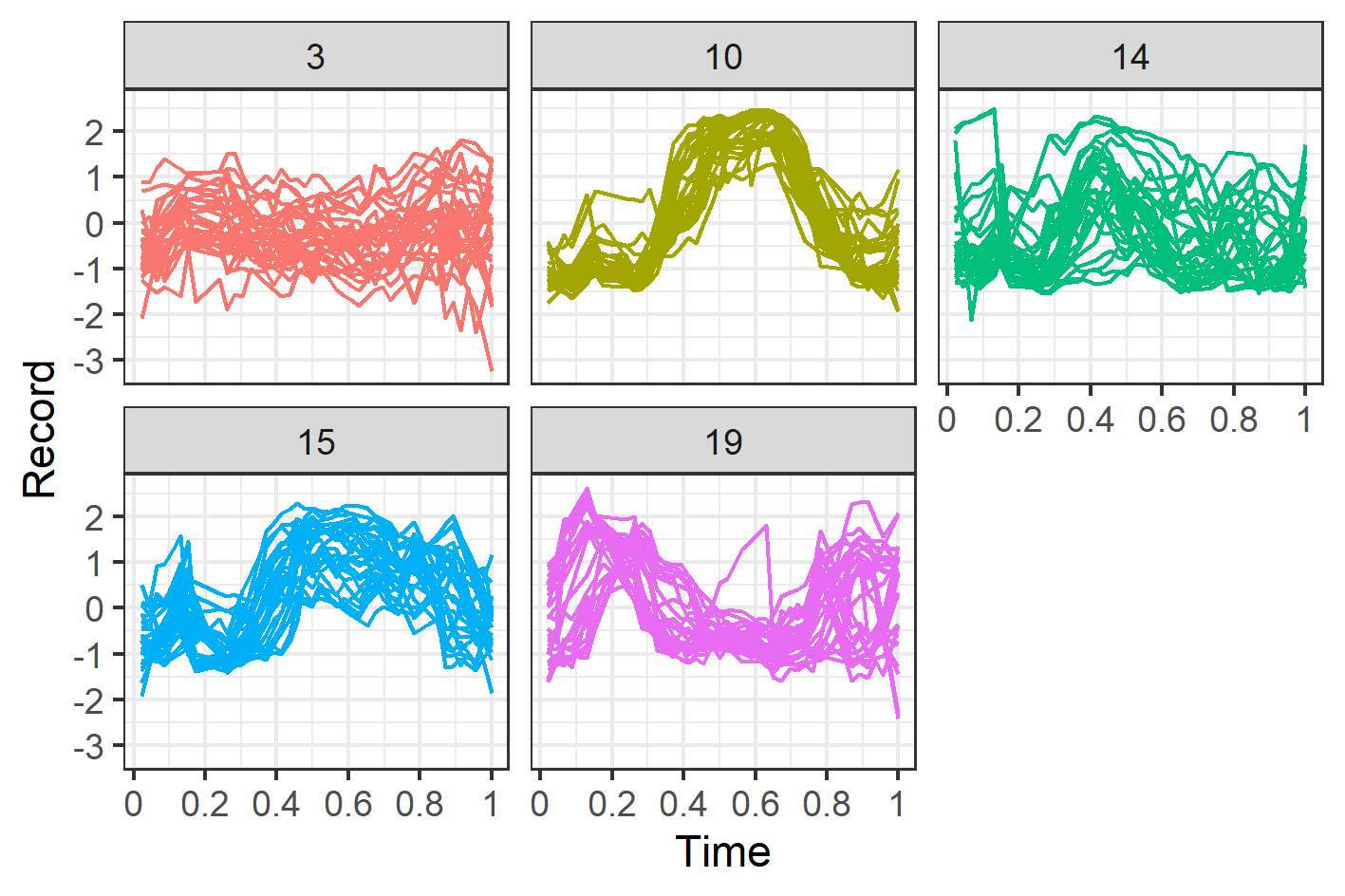}
	\end{center}
	\caption{\label{fig:series2}Time series plots of observations within the five different groups for the crop data.}

\end{figure}

\begin{figure}[h]
	\begin{center}
		\includegraphics[width=110mm]{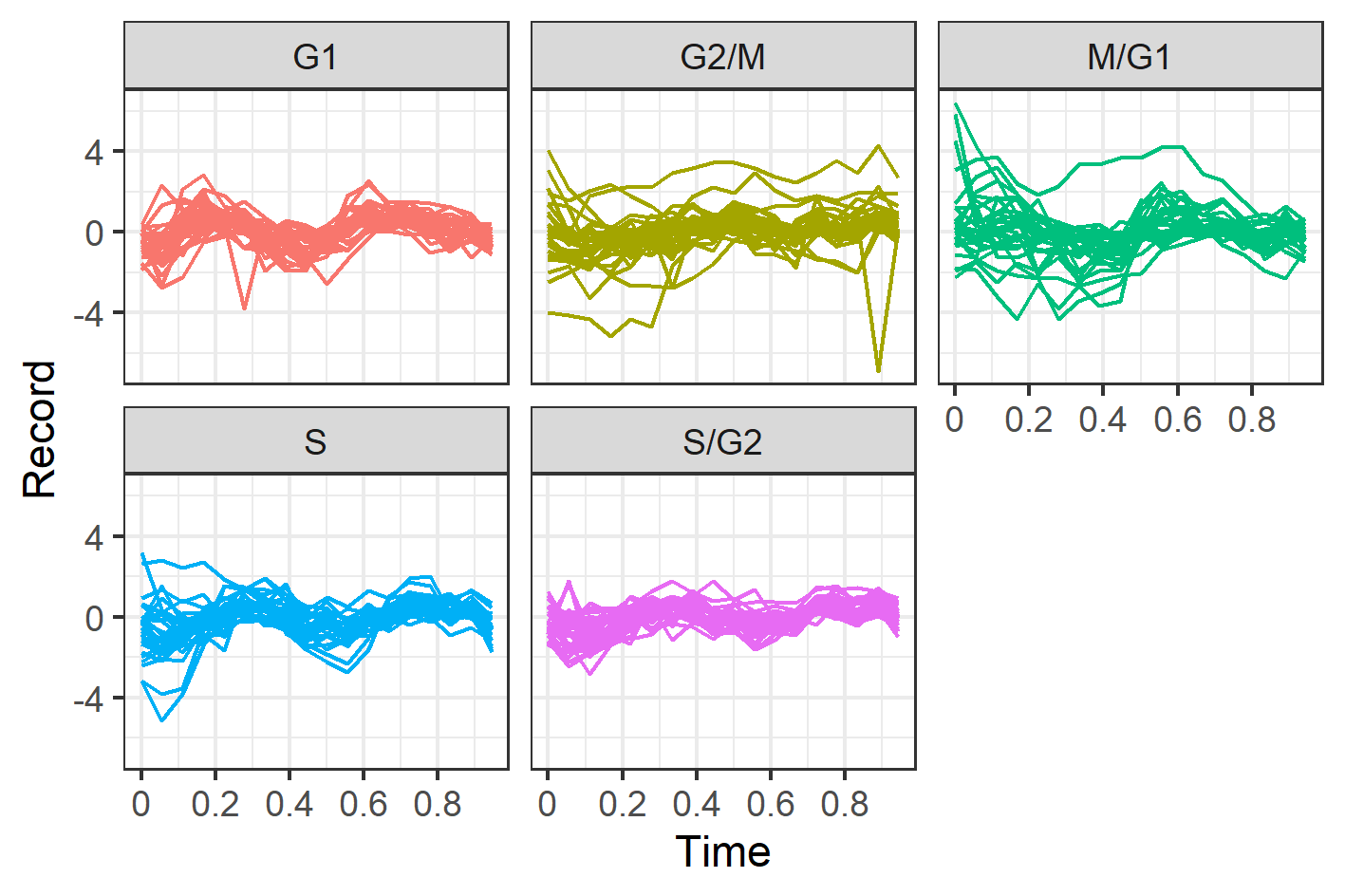}
	\end{center}
	\caption{\label{fig:series3} Time series plots of observations within the five different groups for the DNA data.}
%\end{figure}

%\begin{figure}[h]
	\begin{center}
		\includegraphics[width=110mm]{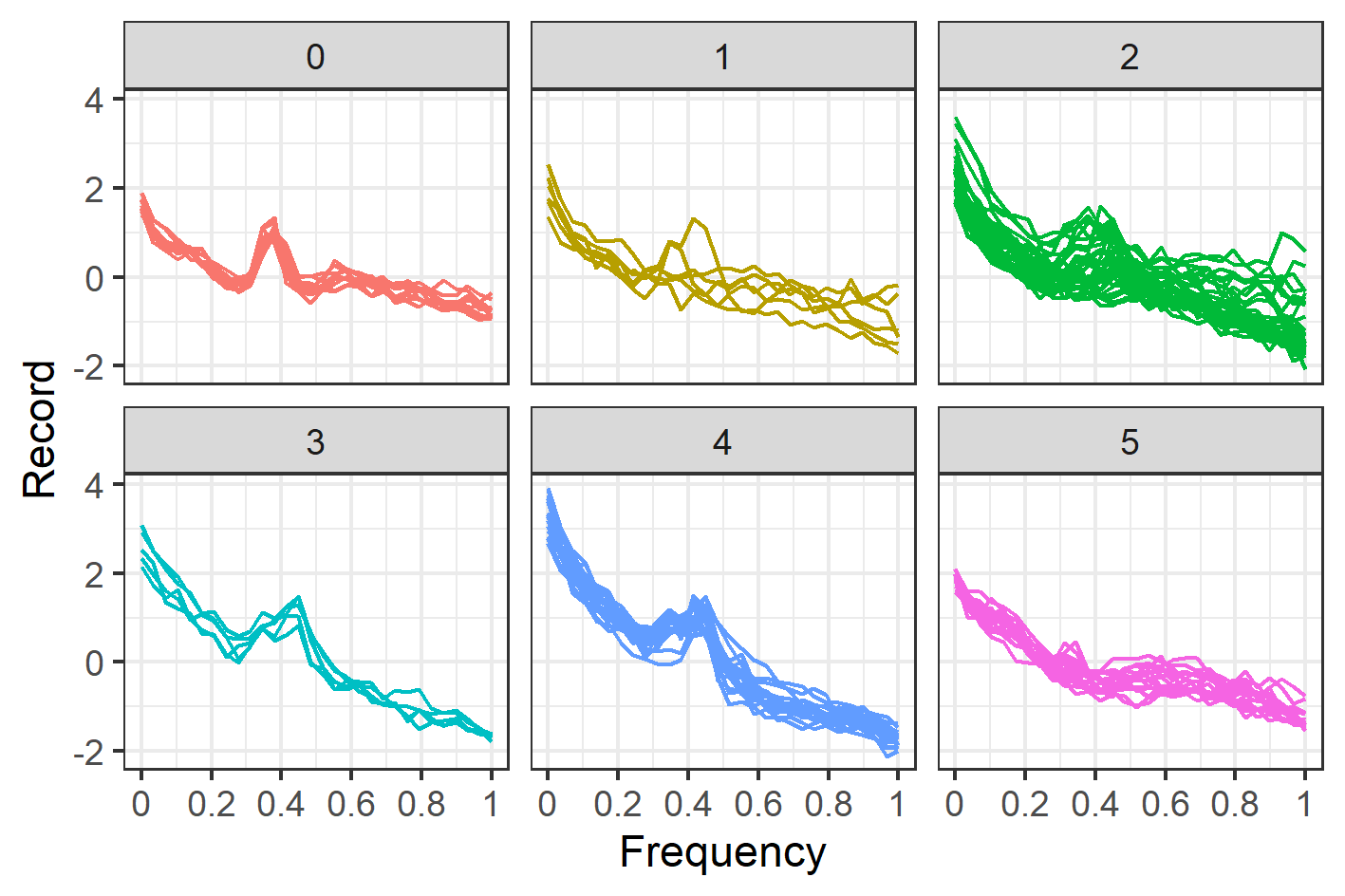}
	\end{center}
	\caption{\label{fig:series4}Time series plots of observations within the 6 different groups for the EEG data. }
	
\end{figure}

\begin{figure}[h]
	\begin{center}
		\includegraphics[width=110mm]{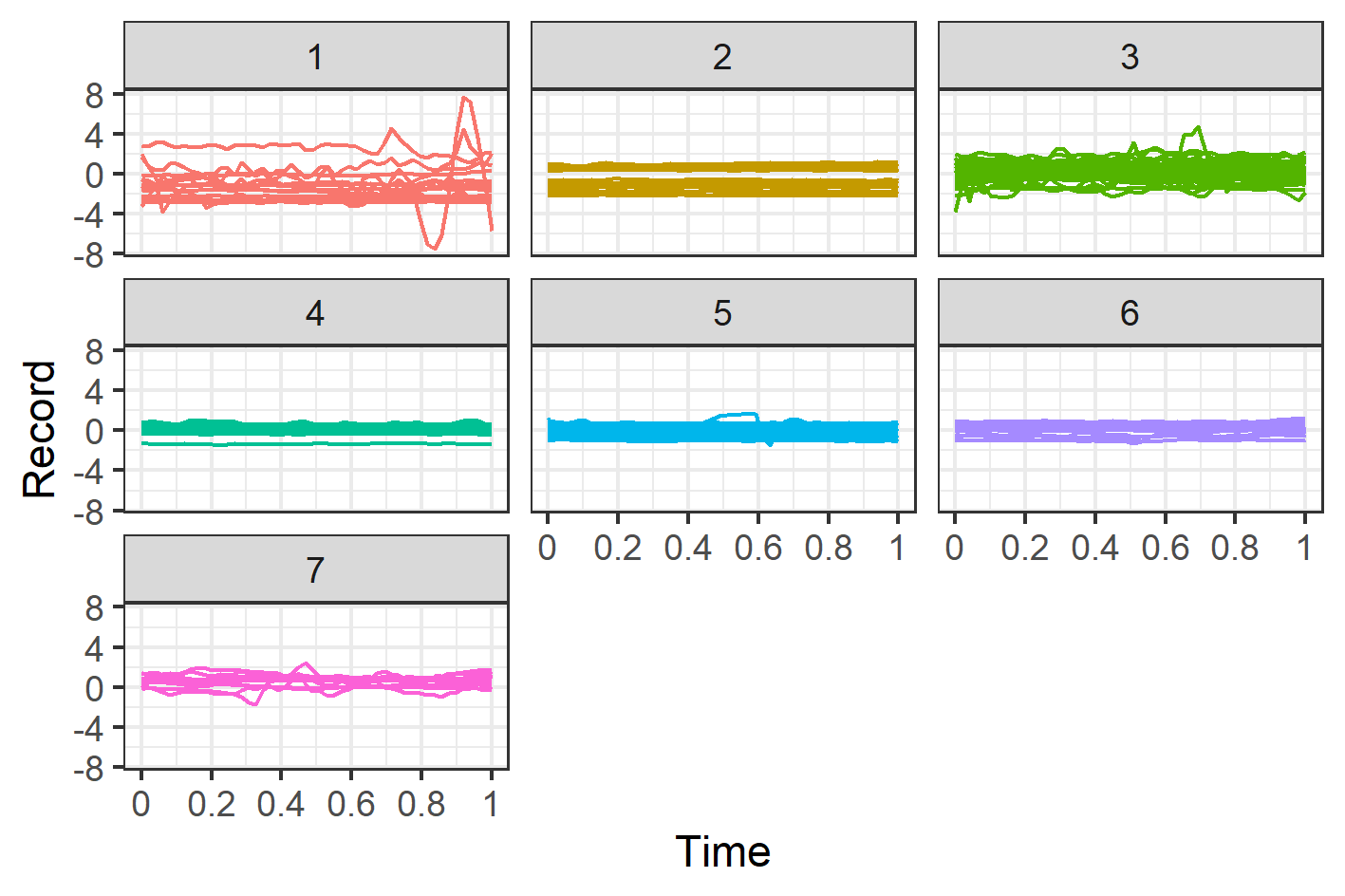}
	\end{center}
	\caption{\label{fig:series5}Time series plots of observations within the 7 different groups for the activity data. }
\end{figure}

\begin{figure}[h]
	\begin{center}
		\includegraphics[width=110mm]{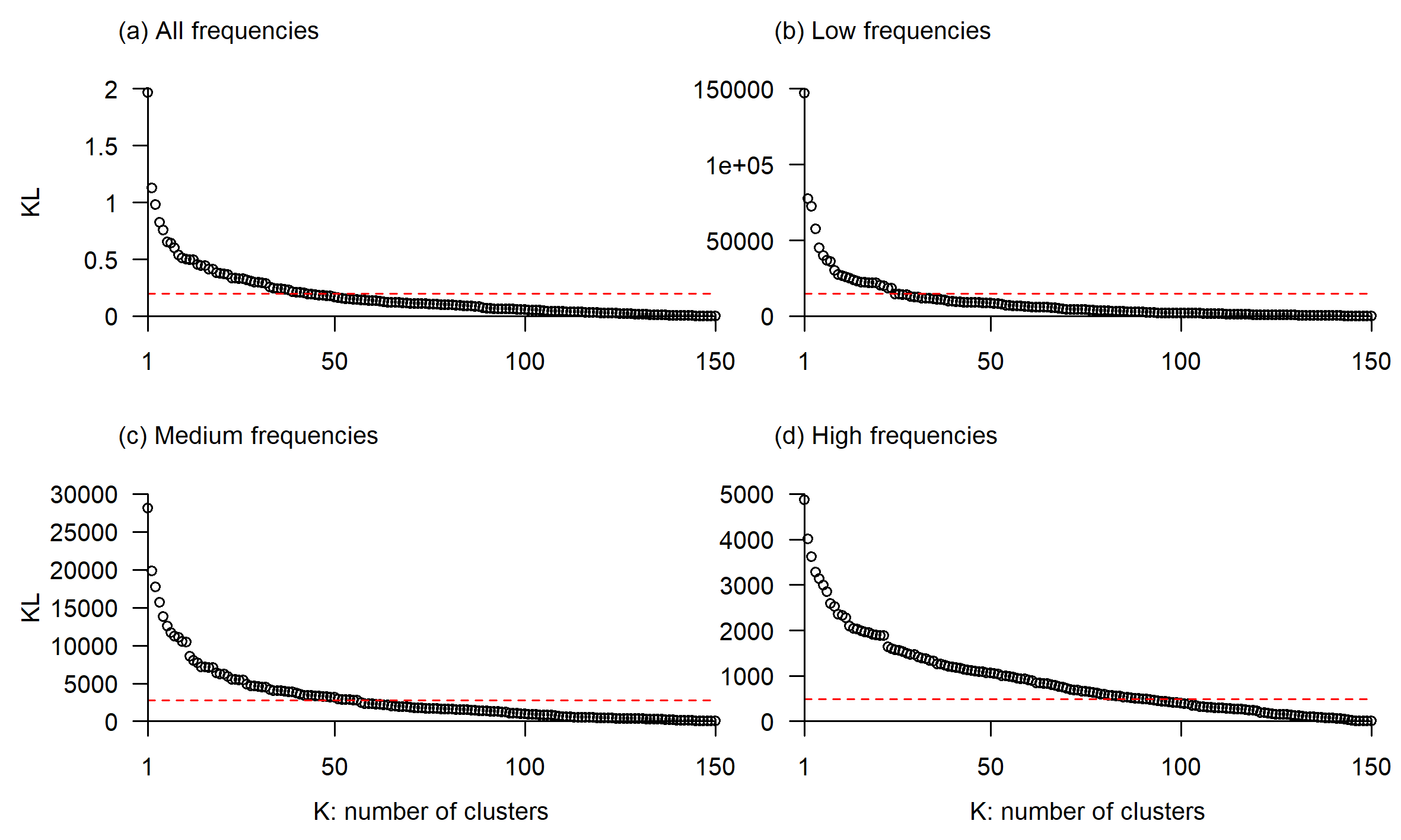}
	\end{center}
	\caption{\label{fig:KL2}Plot of $\text{KL}_K$ versus $K$ for (a) all frequency, (b) low frequency, (c) intermediate frequency and (d) high frequency cases for $Z_{i,A}$ for crop data.  The number of clusters is chosen as the smallest $K$ with $\text{KL}_K$ less than $0.1$}
%\end{figure}

%\begin{figure}[h]
	\begin{center}
		\includegraphics[width=110mm]{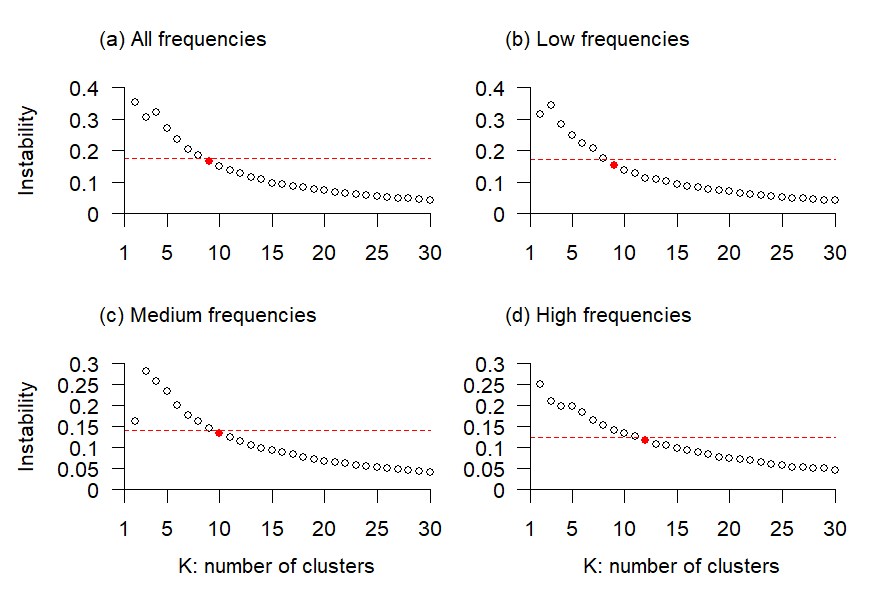}
	\end{center}
	\caption{\label{fig:bootstrap3} Plot of instability $\text{I}_K$ versus $K$ for (a) all frequency, (b) low frequency, (c) intermediate frequency
	and (d) high frequency cases for $Z_{i,A}$ for DNA data.}
\end{figure}	

\begin{figure}[h]
	\begin{center}
		\includegraphics[width=110mm]{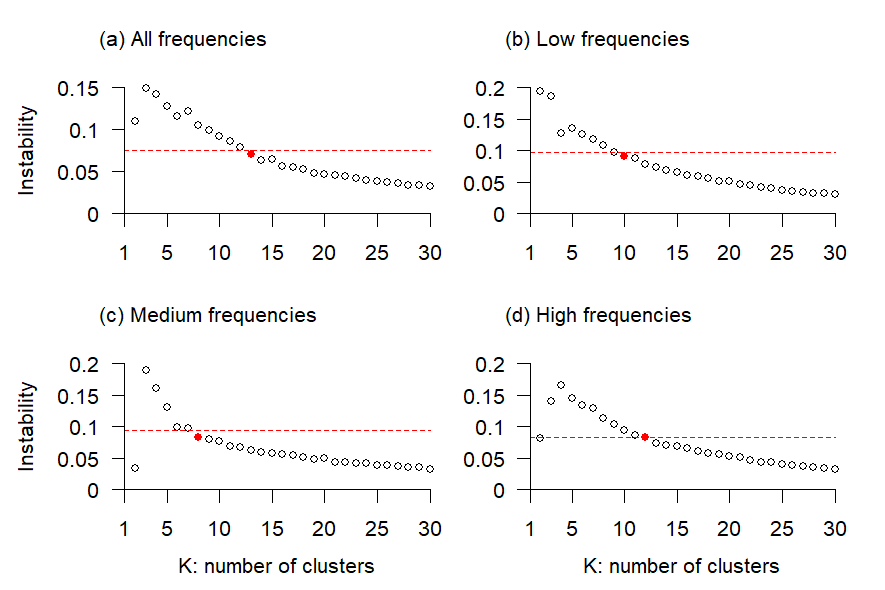}
	\end{center}
	\caption{\label{fig:bootstrap4}Plot of instability $\text{I}_K$ versus $K$ for (a) all frequency, (b) low frequency, (c) intermediate frequency
	and (d) high frequency cases for $Z_{i,A}$ for EEG data.}
%\end{figure}	

%\begin{figure}[h]
\begin{center}
	\includegraphics[width=110mm]{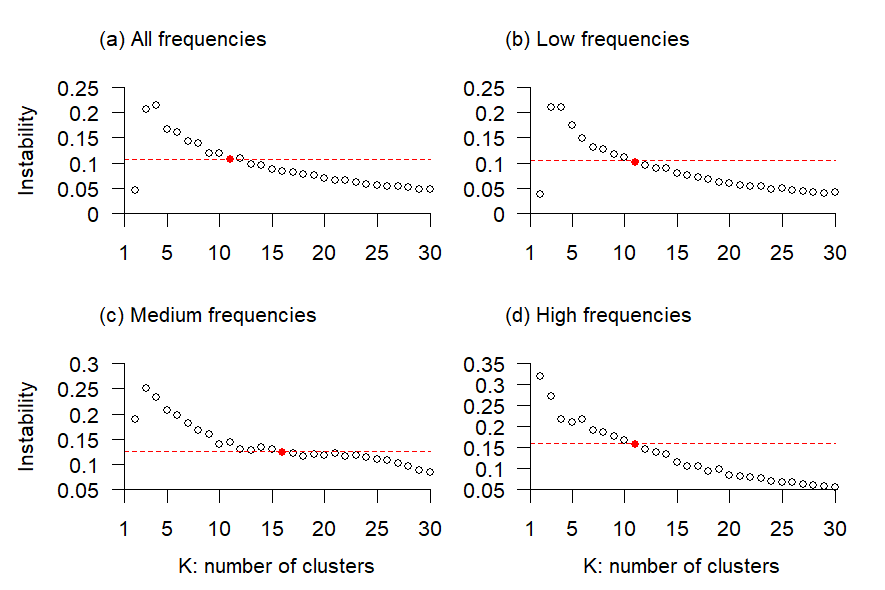}
\end{center}
\caption{\label{fig:bootstrap5}Plot of instability $\text{I}_K$ versus $K$ for (a) all frequency, (b) low frequency, (c) intermediate frequency
	and (d) high frequency cases for $Z_{i,A}$ for accelerometer data.}
\end{figure}	

\begin{figure}[h]
	\begin{center}
		\includegraphics[width=170mm]{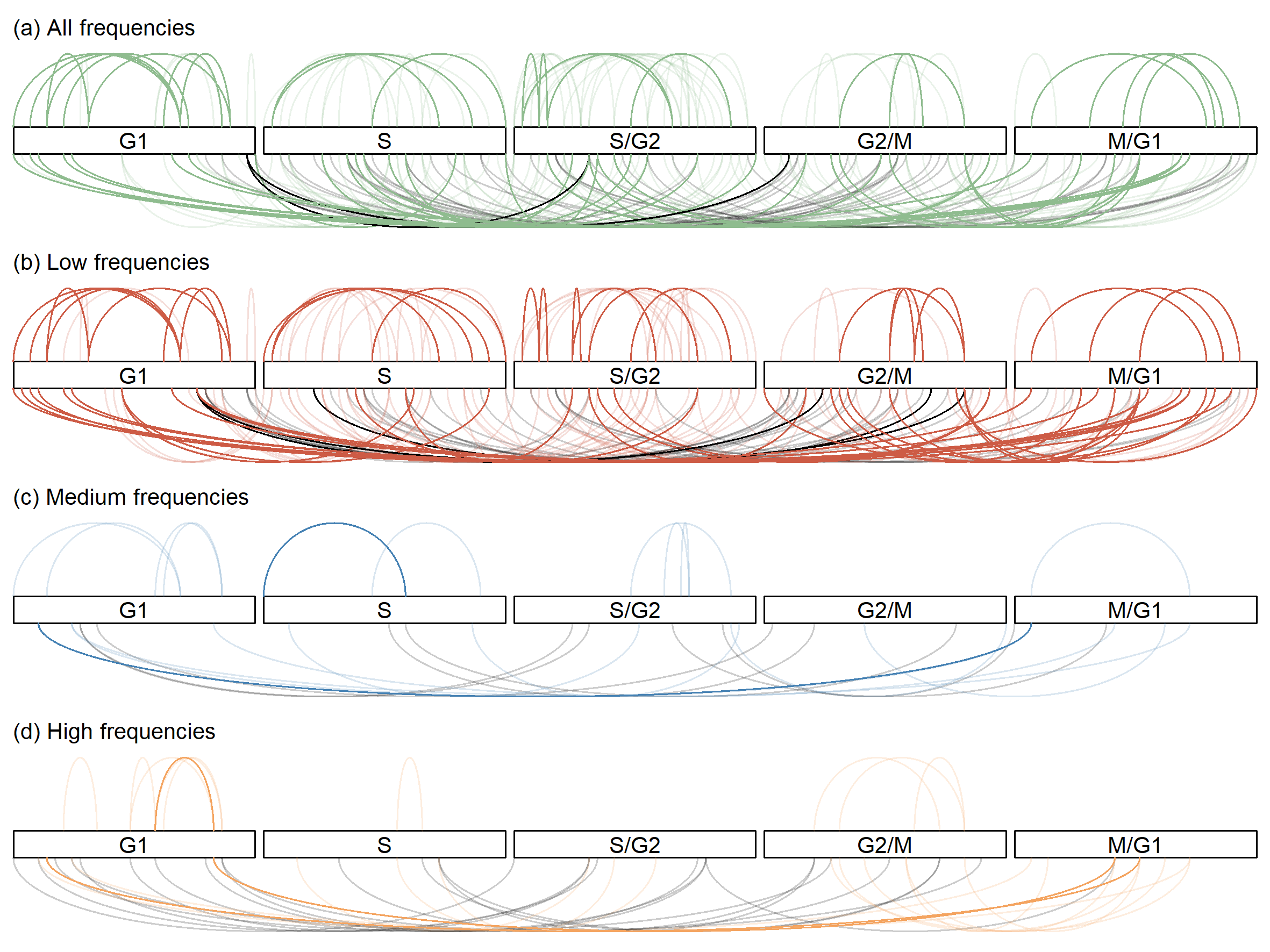}
	\end{center}
	\caption{Pairwise coincidence probabilities for (a) all frequency, (b) low frequency, (c) intermediate frequency and (d) high frequency cases for $Z_{i,A}$ for DNA example.  Interpretation is discussed
	further in the text.  Only 10\% of the links are plotted, randomly chosen.}
	\label{fig:clusterProb3}
\end{figure}	

\begin{figure}[h]
	\begin{center}
		\includegraphics[width=170mm]{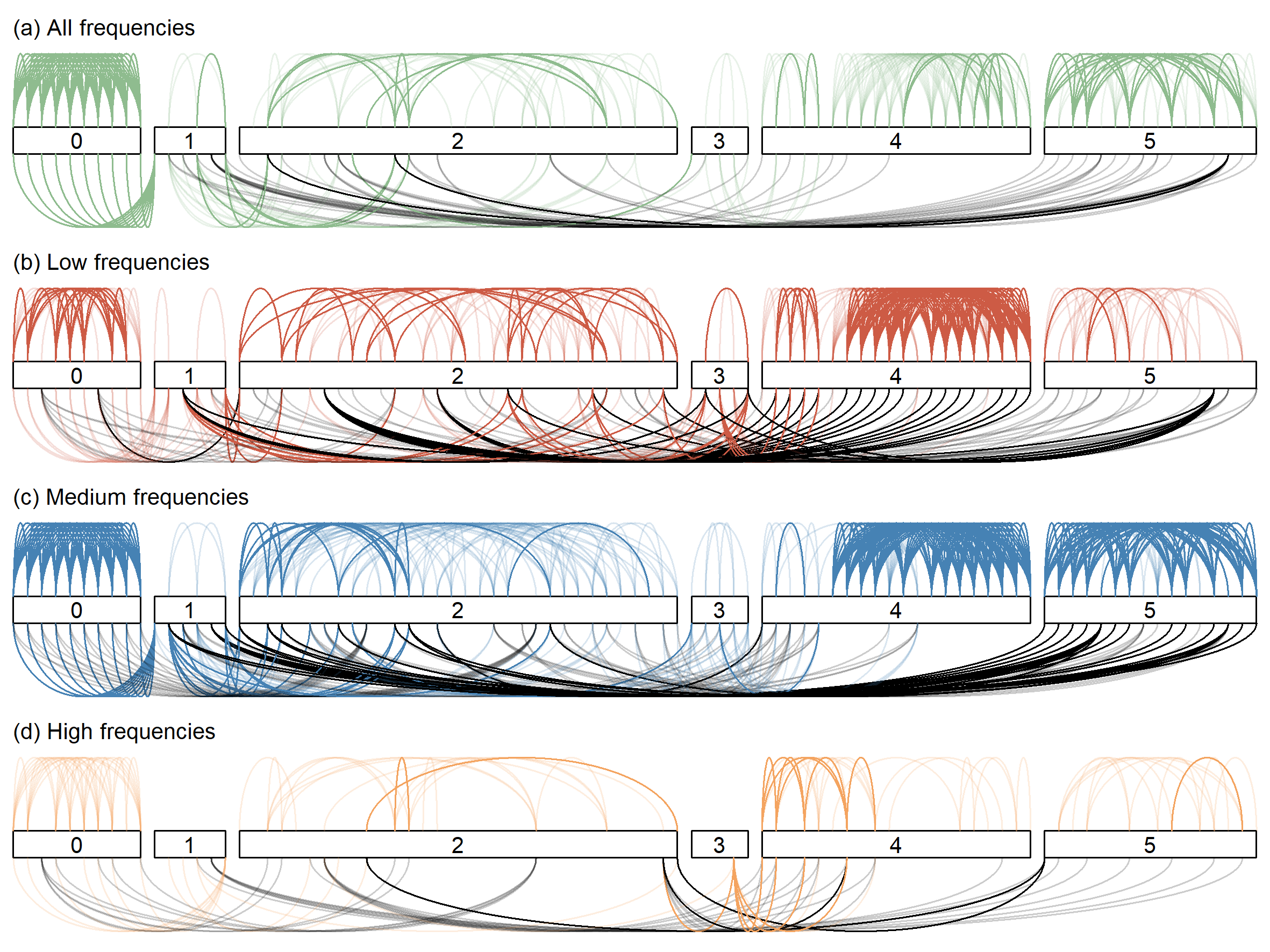}
	\end{center}
	\caption{Pairwise coincidence probabilities for (a) all frequency, (b) low frequency, (c) intermediate frequency and (d) high frequency cases for $Z_{i,A}$ for EEG example.  Interpretation is discussed
	further in the text.}
	\label{fig:clusterProb4}
\end{figure}	

\begin{figure}[h]
	\begin{center}
		\includegraphics[width=170mm]{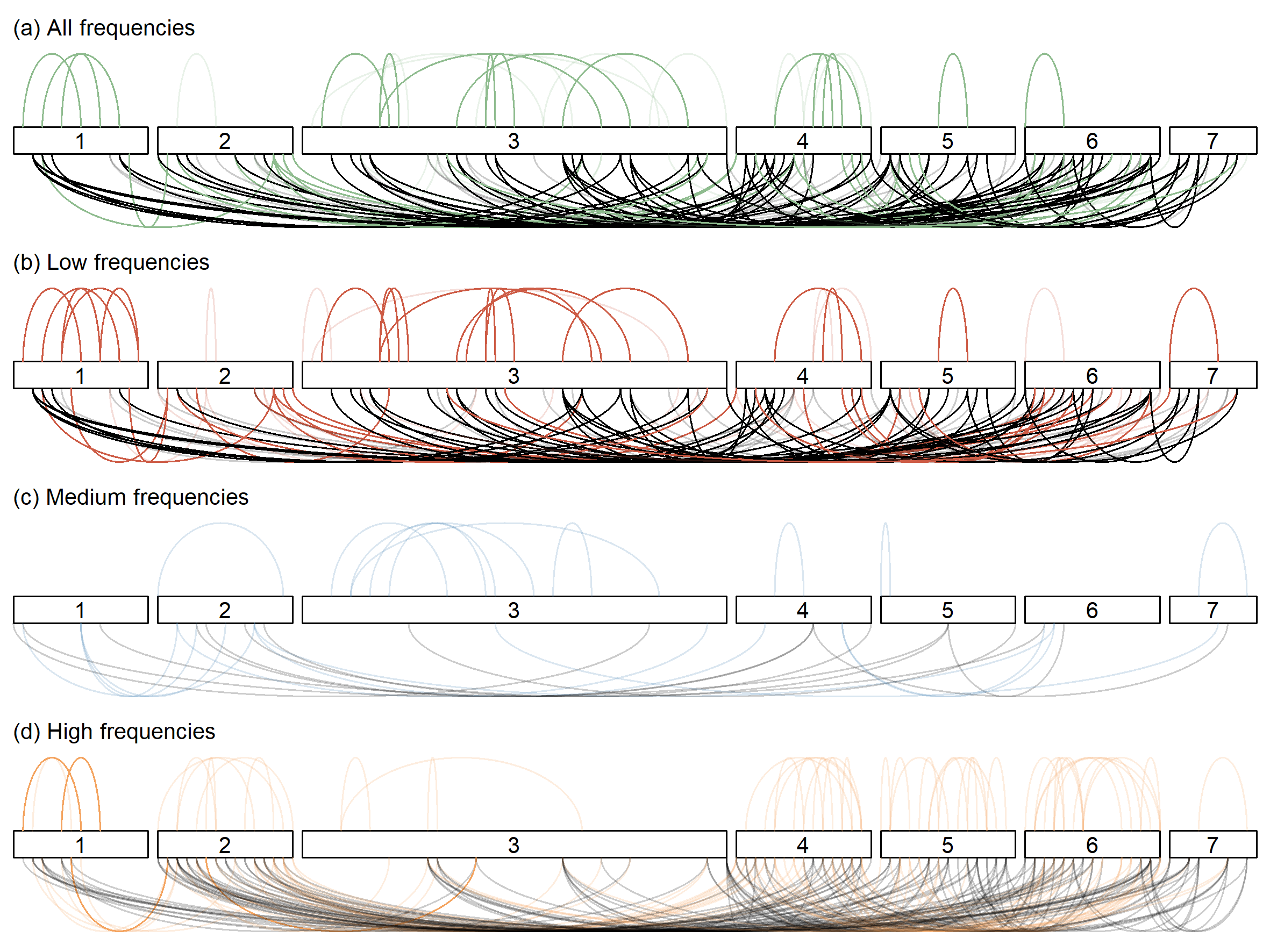}
	\end{center}
	\caption{Pairwise coincidence probabilities for (a) all frequency, (b) low frequency, (c) intermediate frequency and (d) high frequency cases for $Z_{i,A}$ for accelerometer example.  Interpretation is discussed
	further in the text.  Only 10\% of the links are plotted, randomly chosen.}
	\label{fig:clusterProb5}
\end{figure}

\end{document}